\documentclass[prb,twocolumn]{revtex4-1}
\usepackage{amsfonts,amsmath,graphicx,amssymb,bm}
\usepackage[pdftex,plainpages=false,colorlinks=true,linkcolor=blue, citecolor=blue, urlcolor=blue]{hyperref}
\usepackage{natbib}
\usepackage{inputenc}
\usepackage{amsmath,amssymb}
\usepackage{amsbsy}
\usepackage{color,graphicx}
\begin{document}
\title{Antiferromagnetism in the Hubbard model on the honeycomb
  lattice: A two-particle self-consistent study} \date{\today}
\author{S. Arya} \affiliation{The Institute of Mathematical Sciences,
  C.I.T. Campus, Chennai 600 113, India}

\author{P.V. Sriluckshmy} \affiliation{The Institute of Mathematical
  Sciences, C.I.T. Campus, Chennai 600 113, India}

\author{S.R. Hassan} \affiliation{The Institute of Mathematical
  Sciences, C.I.T. Campus, Chennai 600 113, India}

\author{A.-M. S. Tremblay} \affiliation{D\'{e}partement de Physique
  and RQMP, Universit\'{e} de Sherbrooke, Sherbrooke, QC J1K 2R1,
  Canada\\} \affiliation{Canadian Institute for Advanced Research,
  Toronto, Ontario, M5G 1Z8, Canada.\\}

\begin{abstract}
The semimetal to antiferromagnet quantum phase transition of the
Hubbard model on the honeycomb lattice has come to the forefront in
the context of the proposal that a semimetal to spin liquid
transition can occur before the transition to the antiferromagnetic
phase. To study the semimetal to antiferromagnet transition, we
generalize the two-particle self-consistent (TPSC) approach to the
honeycomb lattice (a structure that can be realized in graphene for
example). We show that the critical interaction strength where the
transition occurs is $U_c/t=3.79\pm 0.01$ quite close to the value
$U_c/t=3.869 \pm 0.013$ reported using large-scale quantum Monte Carlo
simulations. This reinforces the conclusion that the semimetal to
spin liquid transition is pre-empted by the transition to the
antiferromagnet. Since TPSC satisfies the Mermin-Wagner theorem, we
 find temperature-dependent results for the antiferromagnetic and ferromagnetic correlation lengths as well as the dependence of double occupancy and of the renormalized spin and charge interactions on the bare interaction strength. We also estimate the value of the crossover temperature to the renormalized classical regime as a function of interaction strength. 
\end{abstract}
\maketitle


\section{INTRODUCTION}

Consider a half-filled band of electrons that interact through a
short-range potential $U$ on a lattice with bandwidth $W$. As one
increases interactions, the ground state can undergo a transition from
a Fermi liquid to a magnetically ordered state for $U$ less than $W$,
but if there is enough frustration, quantum fluctuations may prohibit
long-range order. In that case, upon increasing $U$ further there may be a
Fermi liquid to insulator transition (a Mott transition) where the
insulator, a spin liquid, does not exhibit long-range order.

Spin liquids have been extensively searched for since Anderson's
proposal in the context of high-temperature
superconductors ~\cite{Anderson:1987}. The pyrochlore spin ices with
fractionalized excitations are the best candidates to date for
spin liquid ground states in
three dimensions ~\cite{GingrasReview:2013}. Since quantum fluctuations
are large in low dimension, two-dimensional lattices are especially
good candidates for spin liquid ground states. There is experimental
evidence for such a state of matter in layered organic materials of
the BEDT family that form a highly frustrated triangular
lattice ~\cite{PowellMcKenzieReview:2011}. The first theoretical
proposal for a spin liquid state in the 1970's was in fact for the
triangular lattice ~\cite{Anderson:1973}. Theoretically, evidence for a
spin liquid state has also been found on the kagome
lattice ~\cite{yan_spin-liquid_2011}.


The honeycomb lattice stands as a particularly interesting candidate for a spin liquid ground state because it has the smallest possible coordination for a two-dimensional lattice, leading to large quantum fluctuations. In addition, the Hubbard model on the honeycomb lattice may be relevant for a number of real systems, including graphene, carbon nanotubes, MgB$_2$, etc., as mentioned in Ref.\onlinecite{paiva2005}. 

Much recent work has focused on this model ever since very large scale quantum Monte Carlo simulations made the exciting prediction of a spin liquid over a small range of values of $U$, beyond which antiferromagnetism sets in.~\cite{meng2010}  This claim has been confirmed by further numerical work~\cite{hohenadler_correlation_2011, hohenadler_erratum:_2012,hohenadler_quantum_2012, zheng_particle-hole_2011} but was later disputed by Sorella \emph{et al.}~\cite{sorella2012} using even larger lattices. 

Since methods based on dynamical mean-field theory (DMFT) and its extensions~\cite{Maier:2005,KotliarRMP:2006,LTP:2006}--so-called quantum cluster approaches--are particularly suited to find Mott transitions, they have been used to look for a spin liquid phase between the semimetal and the antiferromagnet. After early single-site DMFT studies ~\cite{tran_finite-temperature_2009,jafari_dynamical_2009,ebrahimkhas_exact_2011}, quantum cluster calculations confirmed the existence of the intermediate spin liquid phase~\cite{wu_quantum_2012,budich_fluctuation-driven_2013,yu_mott_2011} or of a Mott transition~\cite{WuDirac:2010}. However, Hassan \emph{et al.}\cite{hassan2013}, using the cluster dynamical impurity approximation (CDIA) found out that the Mott transition necessary for a spin liquid ground state is in fact pre-empted by antiferromagnetic long-range order. Careful analysis~\cite{LiebschHoneycomb:2011,LiebschWu:2013} of the influence of the cluster shape and of the various implementations of cluster 
extensions of dynamical mean-field theory~\cite{he_cluster_2012,seki_variational_2013} suggest that it is important to apply other quantitative methods to find the precise values of the critical values of $U/W$ for the phase transitions~\cite{Note_1_Arya:2013}. Other approaches that have been applied to this problem are briefly summarized in Refs.~\onlinecite{hohenadler_correlation_2013,LiebschWu:2013}.

Since the Mott transition towards a spin liquid occurs in the absence of long-range order, one can expect that quantum cluster methods give a good upper bound for the occurrence of this transition~\cite{Note_2_Arya:2013}. However, to locate the precise value of $U_c/W$ where antiferromagnetism sets in, it is crucial that the method correctly include long-wavelength quantum fluctuations in the thermodynamic limit. Given that the critical value of $U_c/W$ for antiferromagnetism is of order $2/3$ ~\cite{meng2010}, a semianalytical, nonperturbative technique, valid from weak to intermediate coupling, the two-particle self consistent approach, (TPSC) is especially suited for this problem~\cite{Vilk:1997,TremblayMancini:2011}. This is the approach we use in this paper. Unlike RPA or Hartree-Fock theory, this method satisfies not only conservation laws, but also the Pauli principle, the Mermin-Wagner theorem, and important sum rules for spin and charge fluctuations. TPSC allows us to locate the crossover to the 
renormalized classical regime where the correlation length for antiferromagnetic fluctuations exceeds the thermal de Broglie wavelength. The extrapolation of that crossover line to zero temperature is one of the methods that can be used to find the value of $U_c/W$ where antiferromagnetism sets in. While TPSC has so far been used only in a single-band context, here we generalize it to the two-band case to find $U_c/W$. 

Previous estimates of $U_c$ for the antiferromagnetic transition of the Hubbard model with a nearest-neighbor hopping $t$ on the honeycomb lattice at half-filling and $T=0$ are in the range~\cite{sorella_semi-metal-insulator_1992,martelo:1997,furukawa_antiferromagnetism_2001,paiva2005,sorella2012,meng2010,hohenadler_correlation_2011, hohenadler_erratum:_2012,hohenadler_quantum_2012, zheng_particle-hole_2011,wu_quantum_2012,budich_fluctuation-driven_2013,yu_mott_2011,hassan2013} $3.5t$ to $5t$, much larger than the Hartree-Fock RPA mean-field result~\cite{sorella2012} $2.23t$.  A number of numerical lattice-field theory solutions of the continuum problem (see Ref.~24 of Liebsch and Wu~\cite{LiebschWu:2013}) also suggest an antiferromagnetic phase (more precisely chiral symmetry breaking) at strong coupling. The most accurate estimate for $U_c$ should be the recent large scale quantum Monte Carlo calculation of Sorella \emph{et al.}~\cite{sorella2012}, $U_c/t=3.869\pm 0.013$, close to estimates from high 
temperature 
series expansion~\cite{paiva2005}, $U_{c}~\approx 4t$ and from a projection Monte Carlo method with an optimized initial state by Furukawa \cite{furukawa_antiferromagnetism_2001} $U_{c} \sim 3.6t$. Another accurate recent result, $U_c=3.78t$, is provided by the pinning field approach to quantum Monte Carlo of Assaad and Herbut ~\cite{pinning_assaad_2013}. Other quantum Monte Carlo calculations generally find higher values $U_{c}$. This includes the early ones by Sorella and Tosatti~\cite{sorella_semi-metal-insulator_1992} that yielded $U_{c}=4.5t$, those of Paiva {\emph{et al.}}~\cite{paiva2005} that found $U_{c}~\approx 5t$ and those of Meng {\emph{et al.}}~\cite{meng2010} with $U_{c} > 4.3t$.  The most accurate weak-coupling method that can be compared with TPSC, namely, the functional renormalization group~\cite{honerkamp_density_2008,raghu_topological_2008}, gives $U_{c}~\approx 3.8t$, close to the best estimates mentioned above.


%
%

The paper is organized as follows. In Sec.~\ref{sec:2}, we introduce the model and the notation for the Green function formalism. We generalize the TPSC approach to graphene in Sec.~\ref{sec:3}, obtaining the spin and charge fluctuations with a functional derivative approach. The scaling for the susceptibility is obtained in Sec.~\ref{Sec:Scaling}. The numerical procedure is explained in Sec~\ref{sec:4} and the numerical results are presented in Sec.~\ref{sec:5}. Three appendices contain analytical results that can be obtained for the spin susceptibility.

\section{Model and Green function}\label{sec:2}


The Hamiltonian is given by
\begin{align}
\mathcal{H} &= H_0 + U \sum_i n_{i\uparrow}n_{i\downarrow} \label{graphene} \\
H_0 &= -t \sum_{<ij>\sigma}a_{i\sigma}^{\dagger} b_{j\sigma} + \textrm{H.c.}\label{nih}
\end{align}
where $H_0$ is the noninteracting hopping Hamiltonian. Creation operators for a particle on sublattice $A$ and $B$ are represented by $a^{\dagger}$ and $b^{\dagger}$, respectively, $\sigma$ is the spin of the particle and $<ij>$ represents nearest-neighbor sites on the honeycomb lattice.  Here, $t$ is the hopping parameter and $U$ is the strength of the on-site Coulomb interaction. 

In Fourier space, $H_0$ takes the form
\begin{align}
H_0=
\begin{pmatrix}
	-\mu&-tf(\mathbf{k})\\
	-tf(\mathbf{k})&-\mu
\end{pmatrix}
\end{align}
where 
\begin{align}
f(\mathbf{k})=1 + e^{i \mathbf{k}\cdot \textbf{a}_1} + e^{i\mathbf{k}\cdot \textbf{a}_2}\label{f_k}
\end{align}
with $\textbf{a}_1=\frac{\sqrt{3}}{2}\hat{\mathbf{x}}+\frac{1}{2}\hat{\mathbf{y}}$ and $\textbf{a}_2=\frac{\sqrt{3}}{2}\hat{\mathbf{x}}-\frac{1}{2}\hat{\mathbf{y}}$ the basis vectors of length unity for the underlying triangular Bravais lattice. We take the nearest-neighbor hopping $t$ equal to unity. Similarly, the Planck's constant $\hbar$ and Boltzmann constant $k_B$ are set to unity.

The Green function for the Hamiltonian in Eq.~\eqref{graphene} is a $4$ x $4$ matrix since there are two sublattices and two spin indices.  The Green function matrix $\mathbf{G}$ is diagonal in spin-space because of the spin rotational invariance of the Hamiltonian~\eqref{graphene}. Introducing the notation $1 = (\vec{r_{1}},\tau_{1})$, where $1$ stands for the position on the triangular lattice $\vec{r_{1}}$ and imaginary time $\tau_{1}$, the matrix elements of $\mathbf{G}$ are defined by 
\begin{align}
G_{\alpha\beta}^{\sigma \sigma'}(1,2) & = -\langle T_{\tau} \alpha_{\sigma}(1) \beta_{\sigma}^\dagger(2) \rangle \delta_{\sigma \sigma'},
\label{gf_el}
\end{align}
where $\alpha=a,b$ and $\beta=a,b$ denote sublattice indices and $\sigma,\sigma' = \uparrow, \downarrow$ are the spin indices.

The equation of motion for $G_{\alpha\beta}^{\sigma \sigma'}(1,2)$ in Eq.~\eqref{gf_el} is
\begin{align}
\frac{\partial G_{\alpha\beta}^{\sigma\sigma'}(1,2)}{\partial \tau_1} 
&= -\delta(\tau_1 - \tau_2) \delta_{\mathbf{r}_1\mathbf{r}_2} \delta_{\sigma \sigma'} \delta_{\alpha \beta} \nonumber \\
&- \langle T_{\tau} \frac{\partial }{\partial \tau_1}\alpha_{\sigma}(1) \beta_{\sigma}^\dagger(2) \rangle \delta_{\sigma \sigma'} .
\end{align}
The Heisenberg equation of motion in the grand canonical ensemble yields
\begin{align}
\frac{\partial }{\partial \tau_1}\alpha_{\sigma}(1) & = [\mathcal{H} - \mu \mathcal{N} ,\alpha_{\sigma}(1)] ,
\end{align}
where $\mu$ is the chemical potential and $\mathcal{N}$ is the total-number operator. 
Defining
\begin{align}
h_{\alpha \beta}^{\sigma \sigma'}(1,2) & = -t\sum_{\Delta}\,\delta_{\mathbf{r_{1}}+\Delta,\mathbf{r_{2}}} \, \delta(\tau_1 - \tau_2) \zeta_{\alpha \beta}^{x} \delta_{\sigma \sigma'},
\label{nih}
\end{align}
where $\alpha,\beta = a,b$ are the sublattice indices,  $\Delta$ runs over the nearest neighbors, and $\zeta^{x}$ is the Pauli matrix  
\begin{align}
\zeta^{x} = 
\begin{pmatrix}
0 &1 \\
1 &0
\end{pmatrix},
\end{align}
the equation of motion for the Green function takes the form
\begin{align}
\left( -\frac{\partial }{\partial \tau_1} + \mu\right)&\delta_{\mathbf{r}_1\mathbf{r}_{\bar{3}}} G_{\alpha \beta}^{\sigma \sigma'}(\bar{3},2) - h_{\alpha \eta}^{\sigma \sigma''}(1,\bar{3}) G_{\eta \beta}^{\sigma'' \sigma}(\bar{3},2) \nonumber \\
&= \delta(\tau_1 - \tau_2) \delta_{\mathbf{r}_1\mathbf{r}_2}\delta_{\sigma \sigma'} \delta_{\alpha \beta} \nonumber \\ &- U \langle T_{\tau}  \alpha_{\bar{\sigma}}^{\dagger}(1)\alpha_{\bar{\sigma}}(1)\alpha_{\sigma}(1) \beta_{\sigma}^\dagger(2) \rangle \delta_{\sigma \sigma'},
\end{align}
where a bar over an index like $\bar{3}$ implies summation over the corresponding lattice positions and an integral over imaginary time, while the Einstein summation convention applies to  repeated spin or sublattice indices.

Using 
\begin{align}
\mathbf{G}_0^{-1}(1,2) & = (-\partial_{\tau_1} + \mu)\mathbf{I} - \mathbf{h}(1,2),
\label{nigf}
\end{align}
 for the noninteracting Green function, a short-hand for the above equation of motion is
\begin{align}
\mathbf{G}_{0}^{-1}(1,\bar{3}) \mathbf{G}(\bar{3},2)& = \delta(\tau_1 - \tau_2) \delta_{\mathbf{r}_1\mathbf{r}_2}\mathbf{I} - \mathbf{u}(1,2) 
\end{align}
where the four-point correlation matrix $\mathbf{u}$ is 
\begin{align}
u^{\sigma \sigma'}_{\alpha \beta}(1,2)& = -U\langle T_{\tau}  \alpha_{\bar{\sigma}}^\dagger(1^+)\alpha_{\bar{\sigma}}(1)\alpha_{\sigma}(1) \beta_{\sigma}^\dagger(2) \rangle \delta_{\sigma \sigma'}.\label{u}
\end{align} 

The correlation matrix can be rewritten in terms of the self-energy using Dyson's equation 
\begin{align}
\mathbf{G}^{-1}(1,2) & = \mathbf{G}_{0}^{-1}(1,2) - \boldsymbol{\Sigma}(1,2)
\end{align}
where from the spin-symmetry of the Hamiltonian, the self-energy is block-diagonal in spin subspace. This leads to
\begin{align}
\boldsymbol{\Sigma}(1,\bar{3}) \mathbf{G}  (\bar{3},2) & = \mathbf{u}(1,2).  \label {full_se_pe_tpc}
\end{align}

This clearly shows how the self-energy is related to two-particle correlation functions and to the potential energy in the special case where the first and last indices are equal (with a small positive shift in imaginary time for proper time-order). This well-known relation is obtained without any approximations. This is the multi-band generalization of an important consistency requirement between the self-energy and the double occupancy in the Hubbard model~\cite{Vilk:1997}.

\section{Generalization of TPSC}\label{sec:3}

The two-particle self-consistent (TPSC) approach was developed to study the single-band Hubbard model~\cite{Vilk:1994,Vilk:1996,Vilk:1997,Allen:2003,TremblayMancini:2011}. It has been benchmarked through detailed comparisons with quantum Monte Carlo calculations. This is a nonperturbative method that works best from weak to intermediate values of coupling $U/W$. The key features of this approach are that it satisfies conservation laws, the Pauli principle and the Mermin-Wagner theorem. 

Perturbative methods, which obey conservation laws (like FLEX~\cite{Bickers:1989}), tend to violate the Pauli principle, while those that satisfy the Pauli principle (like the Parquet re-summations) usually violate conservation laws~\cite{Bickers:1991}. Methods like RPA give a finite-temperature transition to an antiferromagnetic state with the long-range order, a scenario prevented by the Mermin-Wagner theorem in two dimensions ~\cite{Mermin:1966}.

Although the spin and charge susceptibilities in TPSC are similar in form to those appearing in RPA, the two methods are fundamentally different. In TPSC, the irreducible spin and charge vertices are not equal. They are assumed to be momentum and frequency independent and are computed self-consistently at the two-particle level in such a way that local sum rules for spin and charge are satisfied. With TPSC, one can study the antiferromagnetic fluctuations in two-dimensional lattices without the unphysical finite-temperature phase transition. The renormalized classical regime, where the fluctuations are large and the correlation length becomes greater than the thermal de Broglie wavelength, can be studied using this theory. This crossover to a renormalized classical regime at a finite temperature is a precursor of the zero-temperature instability to the long-range order. Note, however, that TPSC is not valid deep inside the renormalized classical regime. 

TPSC has been used to demonstrate, for example, that antiferromagnetic fluctuations can induce a pseudogap in two dimensions~\cite{Vilk:1996,Kyung:2004,Hassan:2008} and that $d$-wave superconductivity mediated by these fluctuations is possible~\cite{Kyung:2003}. The method has been generalized to the attractive Hubbard model,~\cite{Allen:2001} and has been extended to the case where one includes a near-neighbor repulsion $V$. This is called extended TPSC, or ETPSC~\cite{davoudi:2006,davoudi:2007,davoudi:2008}.

Here we generalize the method to two bands, but identical atoms within the unit cell. We do not consider the second step of the theory, which gives an improved formula for the self-energy~\cite{Moukouri:2000}. The final form of the theory is very natural but we give a detailed derivation below. The reader may skip to the results section without loss of continuity. The relevant equations are the spin and charge susceptibilities~\eqref{sp_susc} and~\eqref{ch_susc} and the local spin and charge sum rules~\eqref{sr_s} and~\eqref{sr_c} that have to be solved self-consistently with the ansatz~\eqref{A} and~\eqref{sp_vtx}.

\subsection{TPSC ansatz for two bands}

The renormalized interactions for spin and charge can be obtained from functional derivatives of the self-energy $\Sigma$ \cite{Vilk:1997,Allen:2003}.
To obtain $\boldsymbol{\Sigma}$ from its definition in terms of the four-point function $\mathbf{u}$ in Eq.~\eqref{full_se_pe_tpc}, we assume that Hartree-Fock factorization as a product of two-point correlation functions is justified when the four points in $\mathbf{u}$ do not coincide ~\cite{Vilk:1994,Vilk:1997}. But when all four points in $\mathbf{u}$, Eq.~\eqref{u}, are identical, we  impose the exact relation given by 
\begin{align}
\left[ \mathbf{\Sigma}(1,\bar{3})\mathbf{G}(\bar{3},1^+)\right]^{\sigma \sigma'}_{\alpha \beta} \; \delta_{\alpha \beta} \delta_{\sigma \sigma'} = U\langle n_{\alpha \bar{\sigma}}(1) n_{\alpha{\sigma}}(1)\rangle \delta_{\alpha \beta} \delta_{\sigma \sigma'},\label{let}
\end{align}
obtained from Eq.~\eqref{full_se_pe_tpc} when $2\rightarrow1^{+}$ and $\alpha=\beta$, {\em{i.e.}} the positions coincide and the times are such that $\tau_{2} = \tau_{1}+\epsilon$, where $\epsilon$ is positive and infinitesimal. 


Using spin-rotational invariance, since~\eqref{let} is diagonal in spin indices, we focus on the diagonal elements and then,
the above Hartree-Fock-like factorization of Eq.~\eqref{full_se_pe_tpc} can be written as
\begin{align}
\Sigma^\sigma_{\alpha\gamma}(1,\bar{3}) G^\sigma_{\gamma\beta}(\bar{3},2) &= \mathcal{A} G^{\bar{\sigma}}_{\alpha\alpha}(1,1^{+}) G^\sigma_{\alpha\beta}(1,2). \label{hf_like_factor} 
\end{align}
For $2\rightarrow1^{+}$ and $\alpha=\beta$, we find that the exact result ~\eqref{let} is recovered if
\begin{align}
\mathcal{A} & = U \frac{\langle n_{\alpha\bar{\sigma}} n_{\alpha \sigma}\rangle}{\langle n_{\alpha \bar{\sigma}} \rangle \langle n_{\alpha \sigma} \rangle}.  \label{A}
\end{align}
This expression for $\mathcal{A}$ involves 
double occupancy $\langle n_{\alpha \bar{\sigma}} n_{\alpha{\sigma}}\rangle$, which is obtained by the self-consistent calculations explained in the next subsection.

Substituting for $\mathcal{A}$ in Eq.\eqref{hf_like_factor} and right multiplying by $G^{-1}$, we obtain
\begin{align}
\Sigma^\sigma_{\alpha\beta}(1,2)  = U \frac{\langle n_{\alpha\bar{\sigma}} n_{\alpha \sigma}\rangle}{\langle n_{\alpha \bar{\sigma}} \rangle \langle n_{\alpha \sigma} \rangle}  G^{\bar{\sigma}}_{\alpha\alpha}(1,1^{+})\delta(1-2)\delta_{\alpha\beta}. \label{approx_first_se}
\end{align}
This is our first approximation for the self-energy.  It is local and frequency independent. A better approximation can be obtained by including the effects of fluctuations but this is not needed here~\cite{Vilk:1997,Moukouri:2000,TremblayMancini:2011}. As explained in the next section, the functional derivatives of the self-energy obtained above lead to the renormalized vertices for spin and charge. 
\subsection{Spin and charge susceptibilities}
The spin and charge susceptibilities are calculated to reach an understanding of the competing spin and charge ordering transitions in the model. The value of $\mathcal{A}$ in Eq.~\eqref{A} is obtained from these susceptibilities. Unlike RPA, where vertices are the bare $U$ in both spin and charge susceptibilities, in TPSC~\cite{Vilk:1997,TremblayMancini:2011}, spin and charge vertices differ. The renormalized irreducible vertex for spin is denoted by $\mathcal{U}_s$ and for charge by $\mathcal{U}_c$. They will clearly both depend on $U$. 

The spin and charge vertices in the longitudinal spin channel are computed from the local particle-hole irreducible vertices $\Gamma^{\sigma\sigma}$ and $\Gamma^{\sigma \bar{\sigma}}$. These vertices are given by functional derivatives of the self-energy
\begin{align}
\Gamma^{\sigma\sigma'}_{\alpha \beta ,\gamma \zeta}(1,2;3,4) &= \frac{\delta\Sigma^{\sigma}_{\alpha \beta}(1,2)}{\delta G^{\sigma'}_{\gamma \zeta}(3,4)}.
\end{align}
In matrix notation for the sublattice indices, the irreducible spin vertex is given by 
\begin{align}
\mathbf{\Gamma}_{s}(1,2;3,4) &= \frac{\delta\Sigma^{\uparrow}(1,2)}{\delta G^{\downarrow}(3,4)} - \frac{\delta\Sigma^{\uparrow}(1,2)}{\delta G^{\uparrow}(3,4)},
\end{align}
where $\alpha \beta$ corresponds to the row index and $\gamma \zeta$ corresponds to the column index with $\alpha,\, \beta,\,\gamma,\,\zeta = a,b$. From our first approximation for the self-energy, Eq.~\eqref{approx_first_se}, we can calculate these functional derivatives. We check that the functional derivatives of the terms $\frac{\langle n_{\alpha \uparrow} n_{\alpha \downarrow}\rangle}{\langle n_{\alpha \uparrow} \rangle \langle n_{\alpha \downarrow} \rangle}$ cancel by spin-rotational invariance, and we obtain
\begin{align}
\mathbf{\Gamma}_{s}(1,2;3,4) &= \mathcal{U}_s \,\delta(1-3)\delta(1^{+}-4)\delta(1-2),
\end{align}
where the only non-zero elements of $\mathcal{U}_s$ are diagonal in sublattice index and are given by
\begin{align}
\mathcal{U}_s^{\alpha \alpha, \alpha \alpha} &= \mathcal{A} =  U \frac{\langle n_{\alpha\bar{\sigma}} n_{\alpha \sigma}\rangle}{\langle n_{\alpha \bar{\sigma}} \rangle \langle n_{\alpha \sigma} \rangle}.
\label{sp_vtx}
\end{align}

The irreducible charge vertex is 
\begin{align} 
\mathbf{\Gamma}_{c}(1,2;3,4) &= \frac{\delta\Sigma^{\uparrow}(1,2)}{\delta G^{ \downarrow}(3,4)} + \frac{\delta\Sigma^{\uparrow}(1,2)}{\delta G^{\uparrow}(3,4)}.
\end{align}
From the functional derivative of the terms $\frac{\langle n_{\alpha \uparrow} n_{\alpha \downarrow}\rangle}{\langle n_{\alpha \uparrow} \rangle \langle n_{\alpha \downarrow} \rangle}$, we obtain correlation functions of higher order. TPSC makes the assumption that the irreducible charge vertex, like the irreducible spin vertex, is constant and diagonal in sublattice index:
\begin{align}
\mathbf{\Gamma}_{c}(1,2;3,4)&= \mathcal{U}_c\,\delta(1-3)\delta(1^{+}-4)\delta(1-2) .
\end{align}

Introducing the short hand $q = (\mathbf{q},i\nu)$, which stands for the momentum space coordinate $\mathbf{q}$ and the bosonic Matsubara frequency $\nu$, we find a straightforward generalization of the particle-hole Bethe-Salpeter equation \cite{TremblayMancini:2011} to the case of a matrix susceptibility. The corresponding spin susceptibility $\boldsymbol\chi^s$ and the charge susceptibility $\boldsymbol\chi^c $ are given by,
\begin{align}
\boldsymbol\chi^s(q) & = \left(\mathbf{I} - \frac{1}{2}\boldsymbol\chi^0 (q)\mathcal{U}_s\right)^{-1} \boldsymbol\chi^0 (q)\label{sp_susc},\\
\boldsymbol\chi^c(q) & = \left(\mathbf{I} + \frac{1}{2}\boldsymbol\chi^0 (q)\mathcal{U}_c\right)^{-1} \boldsymbol\chi^0 (q)\label{ch_susc},
\end{align}
where $\boldsymbol\chi^0$ is the noninteracting susceptibility (Lindhard function) defined by
\begin{align}
\chi^0_{\alpha\beta,\gamma\zeta}(q) & = -\frac{T}{N^{2}}\sum_{k \sigma \sigma'} {G}^{\sigma \sigma}_{0 \; \gamma\alpha}(k){G}^{\sigma' \sigma'}_{0 \; \beta\zeta}(k+q) \delta_{\sigma \sigma'}. \label{lf}
\end{align}
The summation is over the momentum space $\mathbf{k}$ as well as over the fermionic Matsubara frequencies, and the lattice size is $N\times N$.

The sum rules~\cite{TremblayMancini:2011}  needed for self-consistency are obtained by summing susceptibilities over all momenta and frequencies to recover local equal-time correlation functions. In the spin channel, we find
\begin{align}
\frac{T}{N^{2}} \sum_{q} \chi_{\alpha \alpha, \alpha \alpha}^s(q) & = \langle n_{\alpha \uparrow} \rangle + \langle n_{\alpha \downarrow} \rangle -2\langle n_{\alpha \uparrow} n_{\alpha \downarrow} \rangle.
\label{sr_s}
\end{align}
On the right-hand side, we have used the fact that the Pauli principle must be satisfied in the form $n_{\sigma}^{2} = n_{\sigma}$. The corresponding sum rules for the charge susceptibility are 
\begin{align}
\frac{T}{N^{2}} \sum_{q} \chi_{\alpha \alpha, \alpha \alpha}^c(\mathbf{q}) & = \langle n_{\alpha}^\uparrow \rangle + \langle n_{\alpha}^\downarrow \rangle + 2\langle n_{\alpha}^\uparrow n_{\alpha}^\downarrow \rangle - \langle n_{\alpha}\rangle^2 .
\label{sr_c}
\end{align}

We already have an expression, Eq.~\eqref{sp_vtx}, for $\mathcal{U}_s$ in terms of double occupancy. By substituting this in Eq.~\eqref{sp_susc} for the spin susceptibility, we can evaluate the sum rules given by Eq.~\eqref{sr_s} and obtain the values of the double occupancies $\langle n_{a\uparrow} n_{a\downarrow} \rangle$ and $\langle n_{b\uparrow} n_{b\downarrow} \rangle$, and hence $\mathcal{U}_s$, in a self-consistent manner. By symmetry, here $\langle n_{a\uparrow} n_{a\downarrow} \rangle$ and $\langle n_{b\uparrow} n_{b\downarrow} \rangle$ are equal. We can determine the constant charge vertex $\mathcal{U}_c$ from the sum rules given by Eq.~\eqref{sr_c} once we know the values of double occupancies. Now that we have $\mathcal{U}_s$ and $\mathcal{U}_c$, the susceptibilities can be calculated from Eqs.~\eqref{sp_susc} and~\eqref{ch_susc}.

We can study the fluctuations in the system as a function of temperature $T$ and on-site interaction $U$. The correlation lengths corresponding to various channels in the spin and charge susceptibilities give us an estimate of the magnitudes of the fluctuations and hence let us determine which ordering transition is dominant in the system. The crossover to a renormalized classical regime at lower temperatures can be detected from the corresponding correlation length.

\section{Scaling form for the susceptibilities}\label{Sec:Scaling}

The correlation length is useful to find the renormalized classical regime and the zero-temperature critical value of $U$. In the limit where the correlation lengths are large, a simple analytical form is useful. First, we introduce the notation 
\begin{align}
\chi^{0}_{aa,aa}=\chi^{0}_{aa} ~;~ \chi^{0}_{aa,bb}=\chi^{0}_{ab}~ ;~ \chi^{0}_{bb,aa}=\chi^{0}_{ba}.
\end{align}
Since the $a$ and $b$ sublattices are equivalent, we will use
\begin{align}
\chi^{0}_{aa,aa}=\chi^{0}_{bb,bb}=\chi^{0}_{aa}.
\end{align}
Quite generally, we also have the following equality
\begin{align}
\left(\chi^{0}_{ab}(i\nu)\right)^*=\chi^{0}_{ba}(-i\nu), 
\label{chi_ab-Identity}
\end{align}
where $\nu$ is a bosonic Matsubara frequency.

The spin susceptibilities can conveniently be rewritten in terms of susceptibilities that are either ferromagnetic or antiferromagnetic within a unit cell. First, rewrite the determinant entering the spin susceptibility~(\ref{sp_susc}) as
\begin{widetext} 
\begin{align}
\det\left(\mathbf{I} - \frac{1}{2}\boldsymbol\chi^0 (q)\mathcal{U}_s\right)=
\left[1-\frac{U_s}{2}(\chi^{0}_{aa}(q) - \sqrt{\chi^{0}_{ab}(q)\chi^{0}_{ba}}(q)) \right]\left[1-\frac{U_s}{2}(\chi^{0}_{aa}(q) + \sqrt{\chi^{0}_{ab}(q)\chi^{0}_{ba}}(q)) \right],
\end{align}
\end{widetext}
with an analogous result for the determinant entering the charge susceptibility. Clearly, the location of the poles is determined by the combinations of noninteracting susceptibilities
\begin{align}
\chi^{s,0}_{fm} & = \left(\chi^{0}_{aa} - \sqrt{\chi^{0}_{ab}\chi^{0}_{ba}}\right), \label{fm} \\
\chi^{s,0}_{afm} & = \left(\chi^{0}_{aa} + \sqrt{\chi^{0}_{ab}\chi^{0}_{ba}}\right). \label{afm}
\end{align}
These can be associated with the noninteracting ferromagnetic and antiferromagnetic spin susceptibilities, respectively. Note that the usual definition of antiferromagnetism, which we adopt here, corresponds to alternating spin directions on $a$ and $b$ sublattices but occurs at $\mathbf{q}=0$ as far as wave vectors are concerned. Explicit expressions for the noninteracting susceptibilities appear in Appendix~\ref{non-interacting}. Intraband terms contribute to the ferromagnetic susceptibility while the antiferromagnetic susceptibility involves interband transitions. 

Taking the analogous definition for the interacting case we find, after some algebra detailed in Appendix~\ref{appendix_AFM}, the following scalar equations 
\begin{align}
 \chi^{s}_{fm}(\mathbf{q},i\nu) &= \frac{\chi^{0}_{fm}(\mathbf{q},i\nu)}{1-\frac{U_{s}}{2}\chi^{0}_{fm}(\mathbf{q},i\nu)}\label{fm_simple}\\
\chi^{s}_{afm}(\mathbf{q},i\nu) &= \frac{\chi^{0}_{afm}(\mathbf{q},i\nu)}{1-\frac{U_{s}}{2}\chi^{0}_{afm}(\mathbf{q},i\nu)}
\label{afm_simple}.
\end{align}
They resemble the expressions in the single-band case. Analogous definitions can be made for the charge susceptibilities. 
\begin{figure*}[ht]
	\begin{center}
		\includegraphics[scale = 0.4]{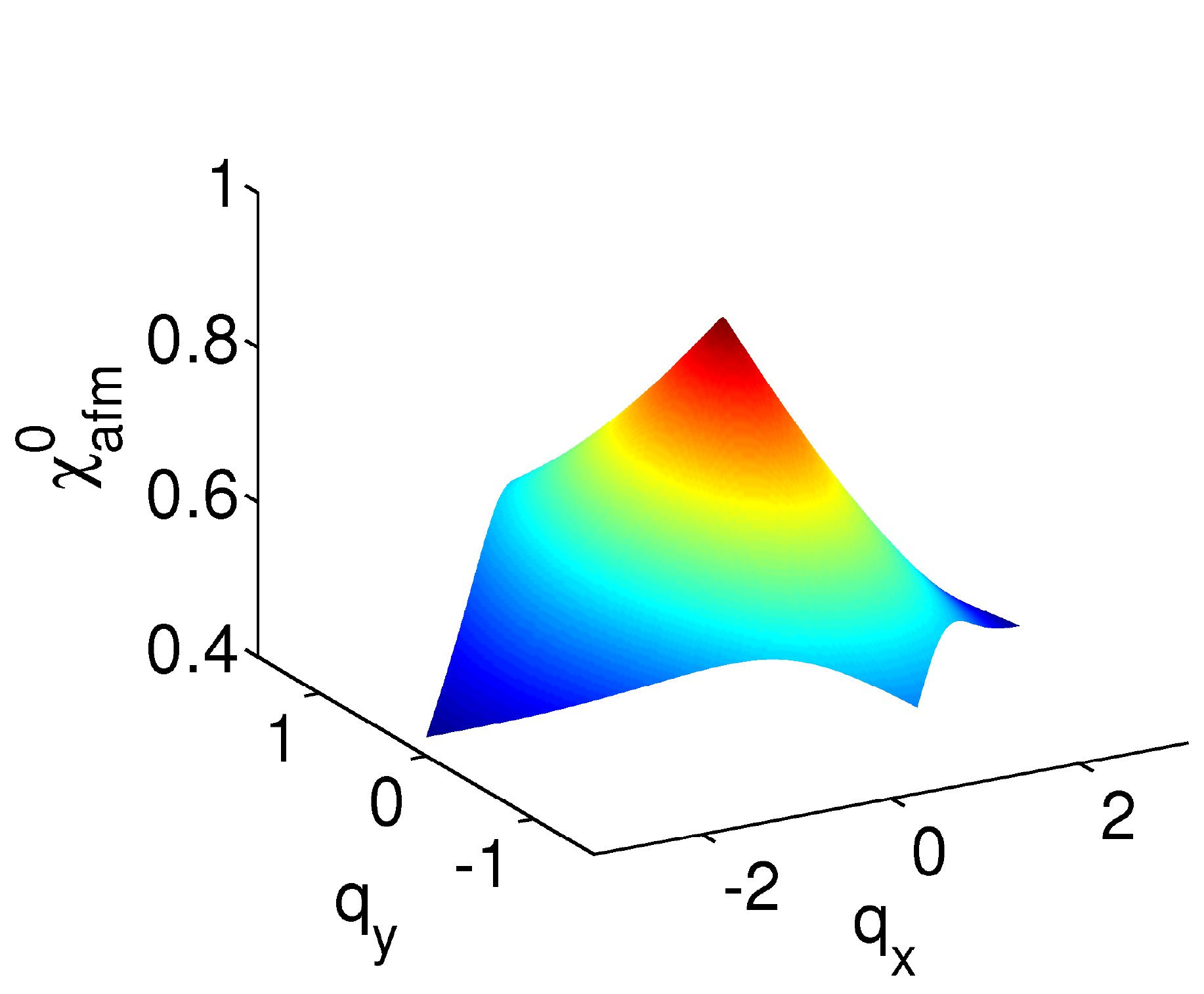}
		\caption{ (Color online) Plot of $\chi_{afm}^{0}(\mathbf{q},i\nu = 0)$ for $T = 0.005$.}
		\label{fig:susc_cone}
	\end{center}
\end{figure*}

The correlation length becomes large when the denominator of the interacting susceptibilities is close to zero at vanishing Matsubara frequency. Taking the antiferromagnetic susceptibility as an example,
in that situation the numerator $\chi^{0}_{afm}(\mathbf{q},i\nu)$ can be replaced by the maximum value $\chi^{0}_{afm}(\mathbf{q} = 0,i\nu=0)$ while in the denominator $\chi^{0}_{afm}(\mathbf{q},i\nu)$ must be expanded about the maximum at $\mathbf{q} = 0, i\nu=0$. 

Because of the Dirac cones, the noninteracting susceptibility in the denominator does not have a derivative at $\mathbf{q} = 0,i\nu=0$. The left derivative and the right derivative as we approach $\mathbf{q} = 0$ are different. In Appendix~\ref{Appendix_analytic_dirac}, we estimate the derivatives in the Dirac approximation. 

We can proceed numerically to confirm the orders of magnitude obtained in Appendix~\ref{Appendix_analytic_dirac}. From the conical shape of the surface plot (see Fig.~\ref{fig:susc_cone}) of the antiferromagnetic susceptibility the dependence is on $q = \sqrt{q_{x}^{2} + q_{y}^{2}}$. The derivative 
is obtained by fitting the data for $\chi^{0}_{afm}(\mathbf{q},0)$ about $\mathbf{q} = 0$, using a form given by $ \chi^{0}_{afm}(\mathbf{q},0) = a + b\,\sqrt{q_{x}^{2}+q_{y}^{2}}+c (q_{x}^{2}+q_{y}^{2})$. The coefficient $b=\partial \chi_{afm}^{0}/\partial q$  is found to be one order of magnitude greater than the coefficient $c$. 

Finally, when the correlation length is large, the above procedure leads to the approximate scaling form for the retarded function
\begin{equation}
\chi_{afm}^{s}(\mathbf{q},\omega+i\delta)= \frac{2 \xi}{U_{s}\xi_{0}} \frac{1}{1 + q\xi + \frac{i\omega \xi}{\Gamma_{0}}} \label{eq:scaling_form}
\end{equation}
where the correlation length is given by
\begin{equation}
\xi = \xi_{0} \frac{U_{s}}{\delta U} .\label{xi(U)}
\end{equation}
In these equations we have used the following definitions: the microscopic length scale
\begin{equation}
 \xi_{0} = -\frac{1}{\chi_{afm}^{0}(\mathbf{q} = 0,i\nu = 0)}\;\frac{\partial \chi_{afm}^{0}(\mathbf{q},i\nu)}{\partial q} \Bigg{|}_{\mathbf{q}=0,i\nu = 0},
 \label{Def_xi0}
\end{equation}
the mean field $U$ for a phase transition
\begin{align}
U_{mf} &= \frac{2}{\chi_{afm}^{0}(\mathbf{q} = 0,i\nu = 0)} ,
\end{align} 
the deviation from the mean field $U$,
\begin{equation}
\delta U = U_{mf} - U_{s},
	\end{equation}
and  
\begin{equation}
\frac{1}{\Gamma_{0}} = \frac{1}{\xi_{0}\chi_{afm}^{0}(\mathbf{q} = 0,i\nu = 0)} \frac{\partial \chi_{afm}^{0''}}{\partial \omega} \Bigg{|}_{\mathbf{q}=0,\omega=0}
\label{Def_Gamma0}
\end{equation}
with $\chi_{afm}^{0''}$ the imaginary part of the retarded susceptibility.

For practical calculations, it is convenient to define the spin correlation lengths for the ferromagnetic and antiferromagnetic channels as
\begin{align}
\xi^{s}_{fm} & = \frac{\chi^{s}_{fm}(\mathbf{q} = \mathbf{0},i\nu = 0)}{\chi^0_{fm}(\mathbf{q} = \mathbf{0},i\nu = 0)} \label{corr_len_fm}, \\ 
\xi^{s}_{afm} & = \frac{\chi^{s}_{afm}(\mathbf{q} = \mathbf{0},i\nu = 0)}{\chi^0_{afm}(\mathbf{q} = \mathbf{0},i\nu = 0)}. \label{corr_len_afm}
\end{align}
Indeed, using the scaling form Eq. \eqref{eq:scaling_form}, the above definition corresponds to
\begin{align}
\xi^{s}_{afm}& =\frac{2 \xi}{U_{s}\xi_{0}}\frac{1}{\chi^0_{fm}(\mathbf{q} = \mathbf{0},i\nu = 0)}\\
&=\frac{\xi}{\xi_{0}}\frac{U_{mf}}{U_s}. \label{xi}
\end{align}
Since $U_{mf}/U_s\sim 1$ when $\xi$ is large, the two definitions of correlation lengths essentially agree in that limit. 

Although similar definitions of correlation lengths can be adopted in the charge channel, these lengths never become large so they are not really useful. 

We end with a note on the critical exponents. TPSC gives us a good estimate of the zero-temperature critical value of $U$, although the exponents usually take values associated with the spherical model~\cite{Dare:1996}.  Accurate values of exponents are usually found with the renormalization group approach.  This is complementary to our approach since the latter methods do not give nonuniversal numbers such as the critical $U$. For graphene, the universality class is that of the Gross-Neveu model~\cite{herbut_interactions_2006,pinning_assaad_2013} with $0.88$ as the value of the correlation length exponent to leading order in $\epsilon$. Instead, we have the value $1$, as follows from Eq.~\eqref{xi(U)}.  From the scaling form~\eqref{eq:scaling_form}, we see that the dynamical critical exponent defined by $\omega\sim \xi^{-z}$ is $z = 1$. Lorentz invariance suggests that $\Gamma_0$ equals the Fermi velocity $v_F$, while a better formula for the $q$ and $\omega$ dependence in the denominator of Eq.~\eqref{eq:
scaling_form} would probably 
replace  $q + \frac{i\omega }{\Gamma_{0}}$ by $\sqrt{q^2-(\omega/v_F)^2}$. Further details appear in Appendix~\ref{Appendix_analytic_dirac}.

\section{Numerical Procedure}\label{sec:4}
We first evaluate the noninteracting susceptibility (Lindhard function) $\boldsymbol\chi^0(q)$ in Eq.~ \eqref{lf}. We then take a guess for $\langle n_{\alpha \uparrow} n_{\alpha \downarrow} \rangle$ to initialize the irreducible spin vertex $\mathcal{U}_s$, Eq.~\eqref{sp_vtx}.  Using Eq.~\eqref{sp_susc}, we compute $\boldsymbol\chi^s$, which, when substituted in the spin sum-rule, Eq.~\eqref{sr_s}, allows us to update the variables $\langle n_{a\uparrow} n_{a\downarrow} \rangle$ and $\langle n_{b\uparrow} n_{b\downarrow} \rangle$ since we know the filling $\langle n_{\alpha \sigma} \rangle = 0.5$ on the right-hand side. We repeat the procedure till we obtain self-consistent solutions for $\langle n_{a\uparrow} n_{a\downarrow} \rangle$ and $\langle n_{b\uparrow} n_{b\downarrow} \rangle$ and thereby obtain the irreducible spin vertex $\mathcal{U}_s$.

A C++ code was written to calculate the noninteracting susceptibilities $\chi^{0}_{aa}$, $\chi^{0}_{ab}$, and $\chi^{0}_{ba}$. FFTs are used in computations to exploit the convolutions in the definitions of the susceptibilities~\cite{Bergeron:2011}. First, the susceptibilities are computed in the position-imaginary time representation where the convolution is just a product. FFT in the position space and a combination of cubic splines and FFT in the imaginary time space are implemented to obtain the final result in the momentum-bosonic Matsubara frequency representation. The real(momentum) space grid is $N \times N$, where $N = 50$, $100$ and $200$ were taken. Since the noninteracting susceptibility obeys, 
\begin{align}
 \frac{T}{N^{2}}\sum_{q}\chi^{0}_{\alpha \alpha}(q) &= \langle n_{\alpha} \rangle = \frac{1}{2}, \label{sum_lf}
\end{align}
we fixed the optimum value for the number of Matsubara frequencies $N_{\omega}$ by requiring that the above be satisfied to $1\%$ accuracy. Accordingly, the range of imaginary time from $0$ to $\beta$ was divided into $N_T = 2 \; N_{\omega}$ slices.

Further comments on finite-size effects and computational procedure may be found at the end of Appendix~\ref{non-interacting}.
\begin{figure}
	\begin{center}
		\includegraphics[width=7cm]{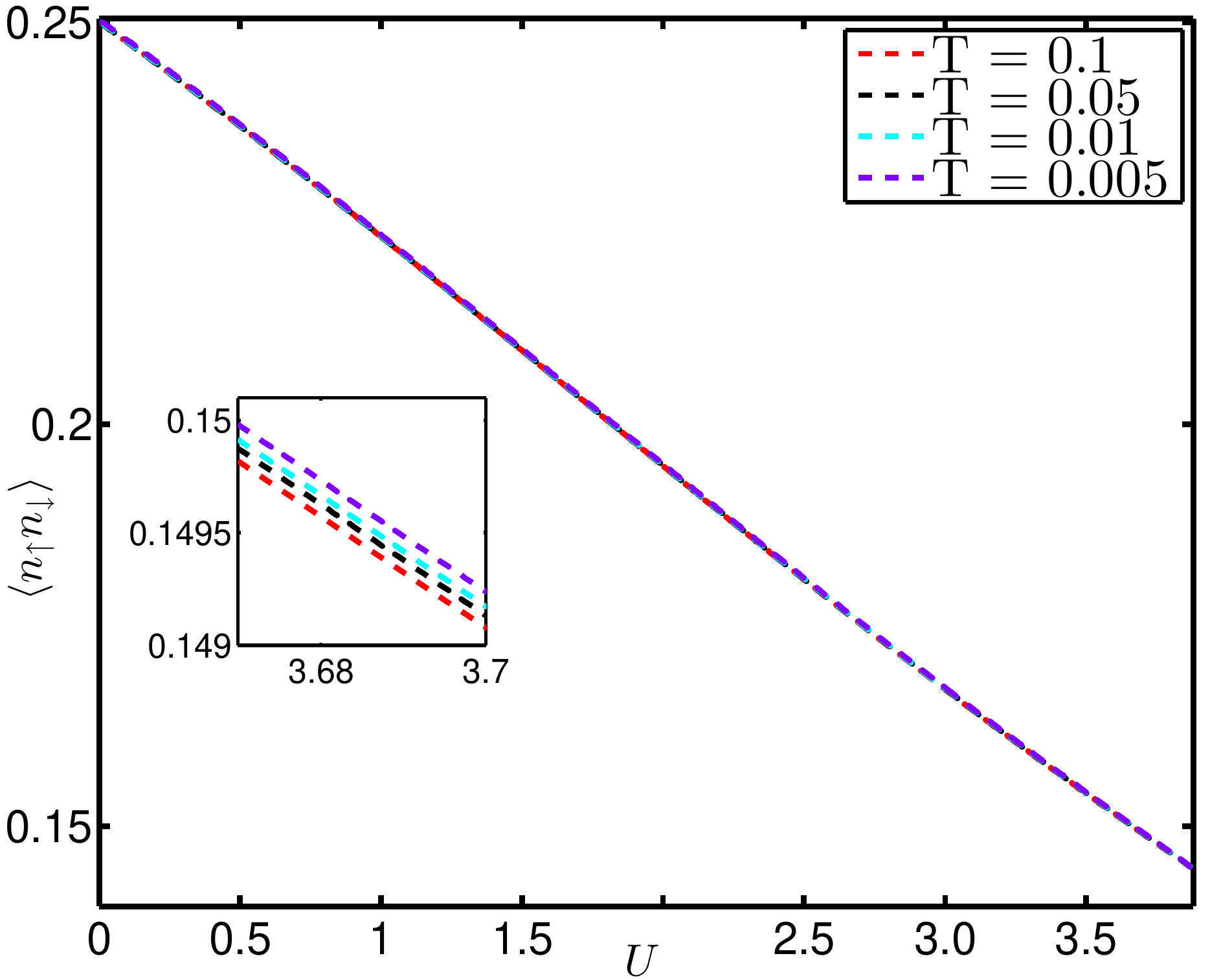}
		\caption{ (Color online) Plot of $\langle n_{\uparrow}n_{\downarrow}\rangle$ as a function of $U$ for given temperatures($N = 100$). The temperature dependence is very small, as can be checked from the inset.}
		\label{fig:nupdown}
	\end{center}
\end{figure}

\begin{figure}[ht]
\begin{center}
\includegraphics[width=7cm]{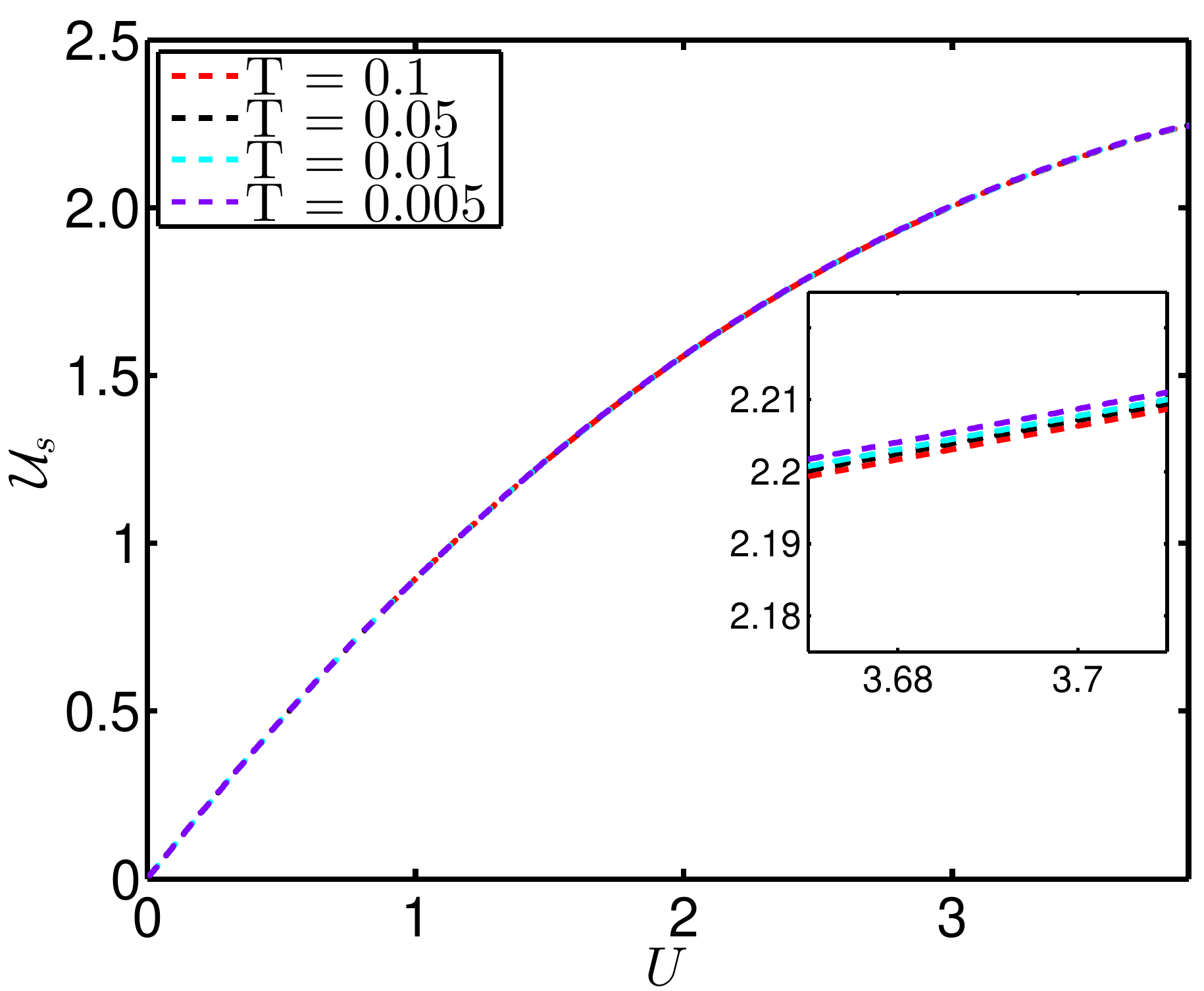}
\caption{ (Color online) Plot of the irreducible vertex for the spin $\mathcal{U}_s$ as a function of $U$ for given temperatures ($N = 100$). The temperature dependence is extremely small as can be seen from the inset.}
\label{fig:Us}
\end{center}
\end{figure}

\section{Results and Discussion}\label{sec:5}
Following the numerical procedure detailed above, we computed the double occupancy $\langle n_{a \uparrow} n_{a \downarrow} \rangle$ self-consistently. This allowed us to obtain the TPSC spin susceptibility, Eq.~\eqref{sp_susc}, as well as the correlation length in the antiferromagnetic channel.

{\em{Double Occupancy}}$\;\langle n_{\uparrow}n_{\downarrow}\rangle$.
Due to bipartite symmetry,  $\langle n_{a\uparrow} n_{a\downarrow} \rangle = \langle n_{b\uparrow} n_{b\downarrow} \rangle$ which we define as $\langle n_{\uparrow}n_{\downarrow}\rangle$. Figure~\ref{fig:nupdown} shows $\langle n_{\uparrow}n_{\downarrow}\rangle$ plotted as a function of interaction $U$ for given temperatures. In the noninteracting case $U = 0$, double occupancy factors into a product of the occupations for up and down electrons. At half-filling, $\langle n_{\uparrow,\downarrow} \rangle =  0.5$, so that $\langle n_{\uparrow}n_{\downarrow}\rangle = \langle n_{\uparrow} \rangle \langle n_{\downarrow} \rangle = 0.25$ for $U = 0$.
As $U$ increases, the energy cost for two electrons occupying a single site increases, thereby leading to a decreasing value of $\langle n_{\uparrow} n_{\downarrow} \rangle$. 

{\em{Spin vertex}} $\mathcal{U}_s$. 
Figure~\ref{fig:Us} shows the spin vertex $\mathcal{U}_s$ as a function of the interaction $U$ for given temperatures.

For very small values of $U$, $\mathcal{U}_s$ is almost equal to $U$. As $U$ increases, $\mathcal{U}_s$ becomes less than $U$ and it shows a tendency to saturate to a constant value. This is a result of Kanamori-Brueckner~\cite{kanamori_electron_1963,Vilk:1994,Vilk:1997} screening: the physics reflects the fact that, as $U$ increases, the two-body wave-function becomes smaller when electrons are on the same site to reduce the probability of double occupancy, thereby decreasing the value of the effective on-site interaction. The maximum energy this can cost is the bandwidth so that at large values of $U$, $\mathcal{U}_s$ saturates to a value of the order of the bandwidth.\cite{Vilk:1997,TremblayMancini:2011}

{\em{Correlation lengths for spin and charge susceptibilities}}. 
With the irreducible spin and charge vertices, we can calculate the spin and charge susceptibilities using the particle-hole Bethe-Salpeter equations \eqref{sp_susc} and \eqref{ch_susc}. From the definitions Eqs.~\eqref{fm} and \eqref{afm} we obtain the spin susceptibilities in the ferromagnetic and antiferromagnetic channels and deduce the correlation lengths in the respective channels from Eqs.~\eqref{corr_len_fm} and \eqref{corr_len_afm}.

The charge correlation lengths are not physically relevant since the irreducible charge vertex is generally larger than $U$ and suppresses the charge susceptibility compared with its noninteracting value, meaning that the correlation lengths are always small and ill-defined. 
\begin{figure}[ht]
\begin{center}
\includegraphics[width=7cm]{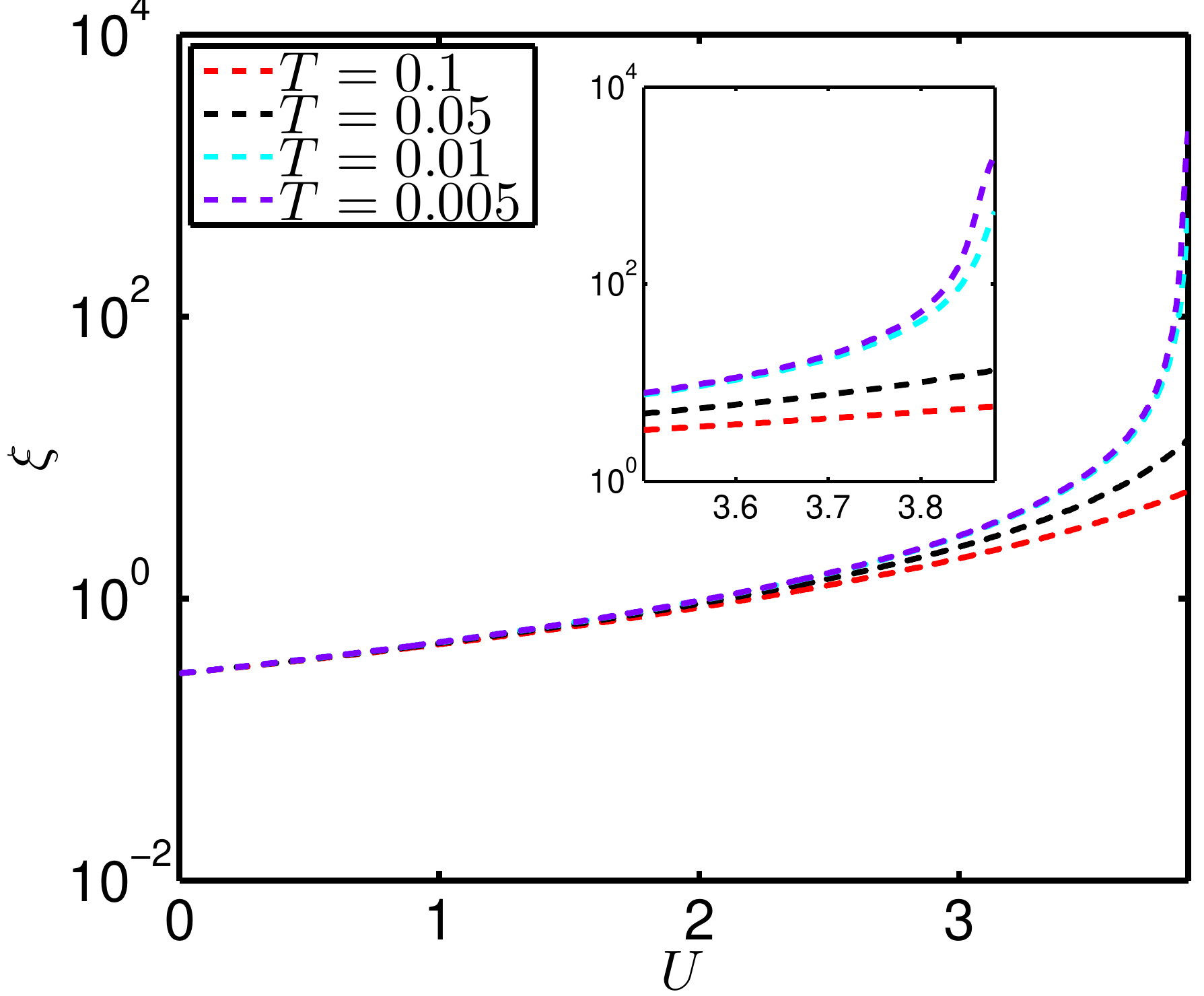}
\caption{ (Color online) Semi-logarithmic plot of the spin correlation length in the antiferromagnetic channel $\xi$ as a function of $U$ for various temperatures. ($N=100$)}
\label{fig:xisafm}
\end{center}
\end{figure}

\begin{figure}[ht]
\begin{center}
\includegraphics[width=7cm]{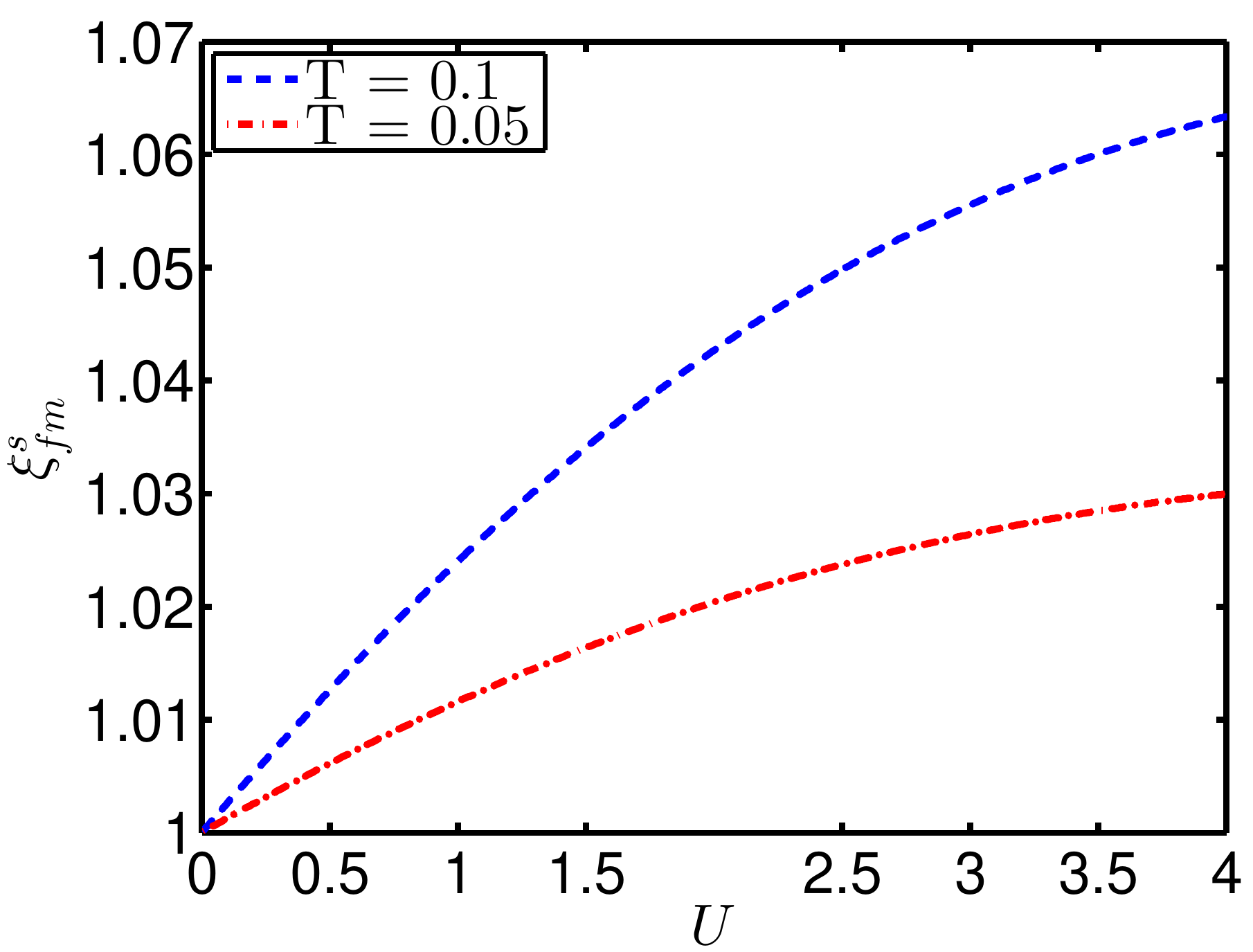}
\caption{ (Color online) Plot of the ferromagnetic correlation length $\xi_{fm}^{s}$ as a function of $U$ for given values of temperature.($N = 100$).}
\label{fig:xisfm}
\end{center}
\end{figure}
Figure~\ref{fig:xisafm} shows the variation of spin correlation length in the antiferromagnetic channel $\xi$ as a function of $U$ for various temperatures. We first obtain the ratio of the interacting susceptibility to the noninteracting susceptibility in the antiferromagnetic channel $\xi_{afm}^{s}$ using Eq. ~\eqref{corr_len_afm} and multiply it by the microscopic length $\xi_{0}$ (Eq.~\eqref{xi}) to obtain the correlation length $\xi$ in units of the lattice spacing. The figure clearly indicates that the spin susceptibility in the antiferromagnetic channel increases steadily with increasing $U$ and with decreasing $T$, as expected, with a clear tendency to diverge at sufficiently large $U$ and low $T$. The quantitative accuracy of the results cannot be trusted for correlation lengths smaller than unity or larger than about half the system size.    

Figure~\ref{fig:xisfm} shows the plots for the ferromagnetic correlation length $\xi_{fm}^{s}$, estimated from the ratio of the interacting susceptibility to the noninteracting susceptibility, Eq.~\eqref{corr_len_fm},  as a function of $U$ for various temperatures for $N = 100$. We can see that the ratio decreases as temperature decreases. Thus in the ferromagnetic channel, the correlation length never becomes larger than the lattice spacing and hence we do not focus on that case. 

\begin{figure}[ht]
	\begin{center}
		\includegraphics[width=8cm]{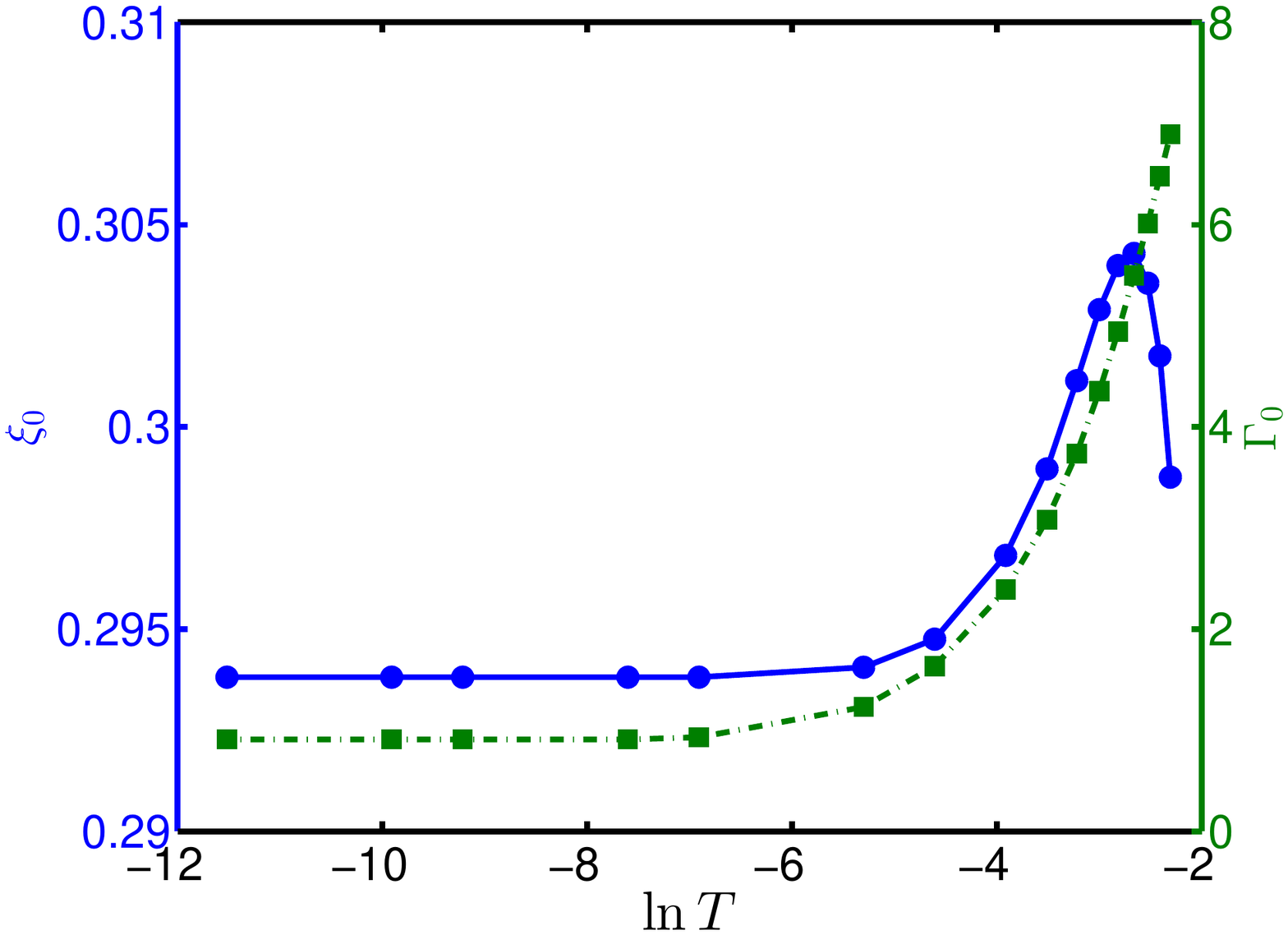}
		\caption{ (Color online) Plot of $\xi_{0}$ (dashed blue line with circles) and of $\Gamma_{0}$ (dot-dashed green line with squares) as a function of $\ln T$ for the range $T = 0.001$ to $T = 0.1$. Each quantity has its own vertical axis: left for $\xi_{0}$ and right for $\Gamma_{0}$.}
		\label{fig:xi0_G0inv}
	\end{center}
\end{figure}

{\em{$\xi_{0}$ and $\Gamma_{0}^{-1}$}}. We numerically determine $\xi_{0}$ and $\Gamma_{0}$ for the spin susceptibility in the antiferromagnetic channel from the relations given in Eqs.~\eqref{Def_xi0} and \eqref{Def_Gamma0} respectively. Figure~\ref{fig:xi0_G0inv} shows the resulting temperature dependence of $\Gamma_{0}$ and of $\xi_{0}$. The microscopic length scale $\xi_{0}$ is almost $T$ independent and of order 0.25. $\Gamma_{0}$ converges to the expected value $v_F = \frac{\sqrt{3}}{2}$ only at very low $T$ and for large values of $N \sim 2000$. Further discussion and numerical estimates appear in Appendix~\ref{Appendix_analytic_dirac}. 

\begin{figure}[ht]
	\begin{center}
		\includegraphics[width=8cm]{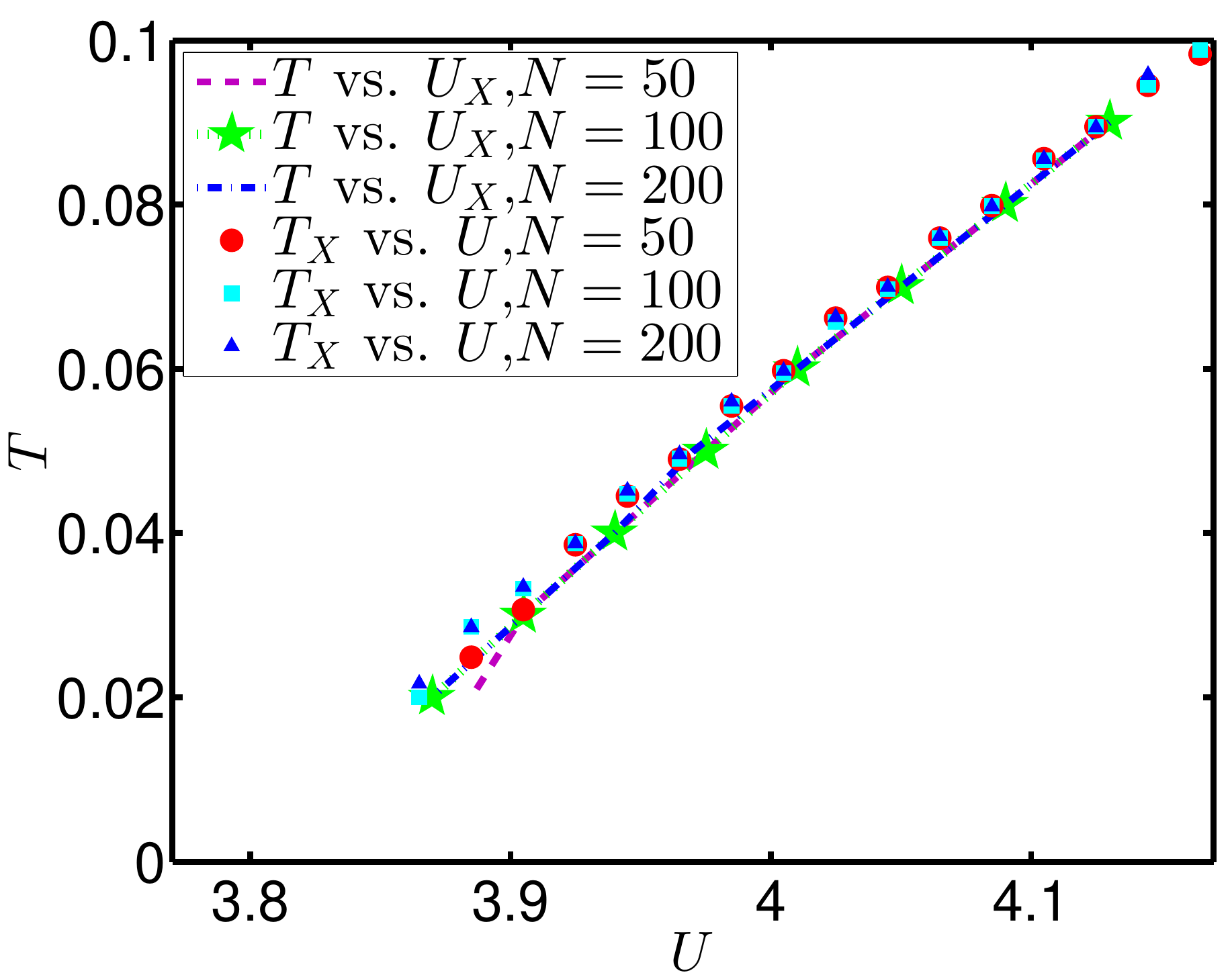}
		\caption{ (Color online) Plots of the crossover temperature as a function of interaction strength for given values of $N$. The crossover temperature was determined from $\xi =  {v_{F}}/{T}$ scanning $U$ at fixed $T$, and scanning $T$ at fixed $U$ to obtain an estimate of the error in finding the intersection $\xi(U,T) =  {v_{F}}/{T}$. }
		\label{fig:crossover}
	\end{center}
\end{figure}

\begin{figure}[ht]
\begin{center}
\includegraphics[width=7cm]{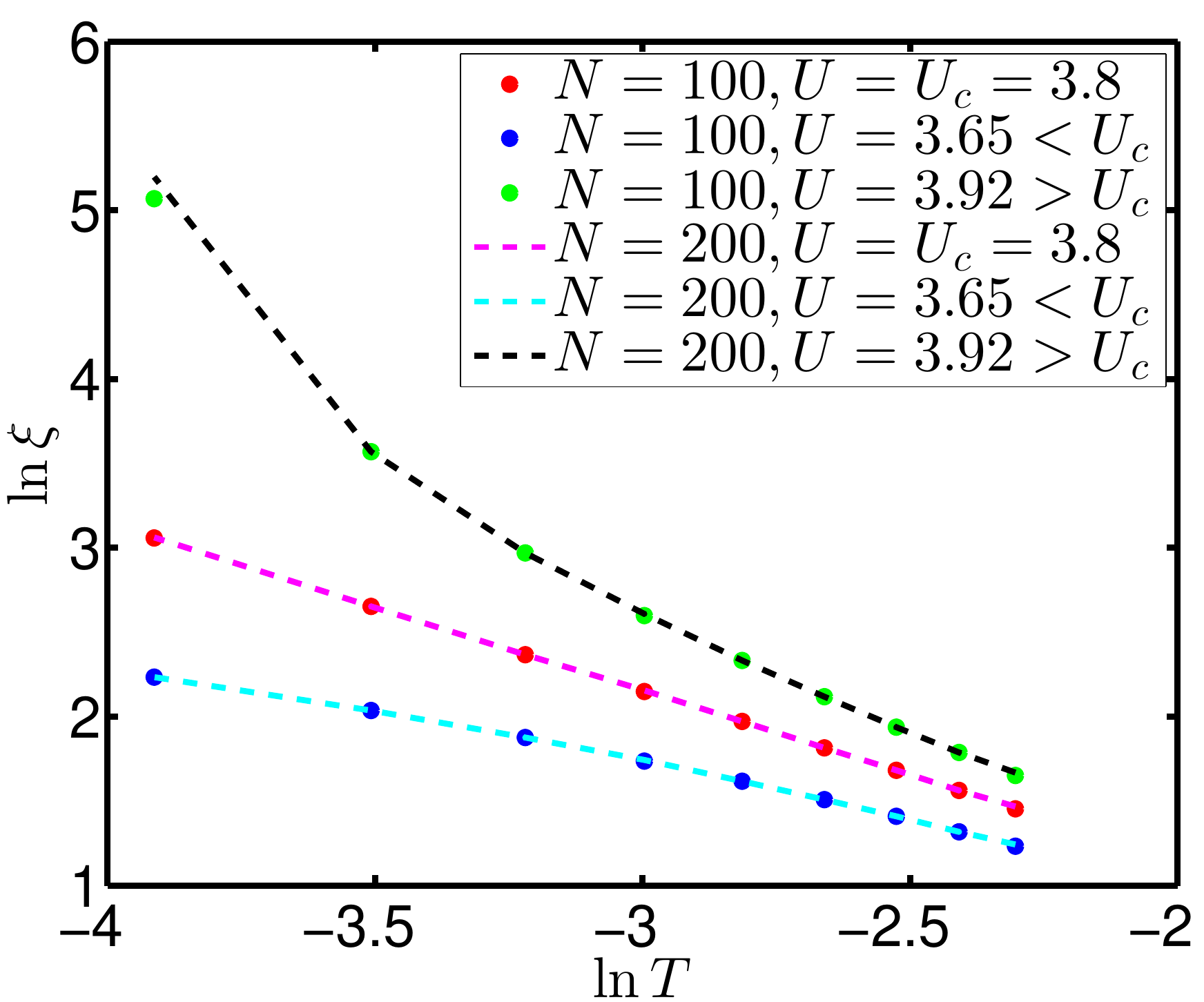}
\caption{ (Color online) Plots of $\ln \xi$ as a function of $\ln T$ for various values of $U$ ($N = 100$ and $200$). The straight dashed magenta line corresponds to a pure power law, $z=1$, hence to the value $U_c$ for the quantum critical point.}
\label{fig:log_xi_log_T_U}
\end{center}
\end{figure}

{\em{Crossover temperature and $U_{c}$}}. For $U>U_c$ there is antiferromagnetism at $T=0$. We expect then that, when $U>U_c$, below a crossover temperature $T_{X}$, the antiferromagnetic correlation length becomes so large that one enters the renormalized-classical regime where the characteristic spin fluctuation frequency $\omega_{sf}$ is less than temperature. And indeed, since the scaling form~\eqref{eq:scaling_form} implies that $\omega_{sf}\sim \xi^{-1}$  and $\xi$ increases faster than $T$ at sufficiently low $T$ when $U$ is larger than $U_c$ (Fig.~\ref{fig:log_xi_log_T_U}), this implies that for $U>U_c$ there is necessarily a temperature below which the condition $\omega_{sf}<T$ is realized.  In the more standard case where the dynamical critical exponent satisfies $z=2$, a pseudogap in the single-particle density of states appears at a temperature smaller than that where $\omega_{sf}\sim T$.  At that temperature,  $\xi$  becomes larger than the thermal de Broglie wavelength $v_F/T$ (with $v_F$ the 
Fermi velocity).~\cite{Vilk:1995,Vilk:1997,TremblayMancini:2011} The question of the appearance of a pseudogap in the present case remains to be investigated, but it is expected as a precursor since there is a real gap in the antiferromagnetic state.  The pseudogap should appear basically when we enter the renormalized classical regime since here frequency and wavevector scale in the same way. 

We thus define the crossover temperature to the renormalized classical regime by the condition $\xi =  {v_{F}}/{T}$, with $v_F$ at the Dirac point. In order to extract the crossover temperature for a fixed value of $U$, we plot the correlation length as a function of temperature and pick the value of temperature ($T_{X}$) where this plot intersects the plot of $v_{F}/T$ as a function of temperature. Similarly, for a fixed value of $T$, we can pick the value of interaction ($U_{X}$) where the correlation length exceeds ${v_{F}}/{T}$. Figure~\ref{fig:crossover} shows the plots of crossover temperature as a function of interaction determined using both approaches detailed above, for $N = 50$, $100$, and $200$. By quadratic and linear extrapolations of the curves to zero temperature, one obtains the results for $U_c$ that appear in Table \ref{table:crossover}.

\begin{table}
	\begin{center}
		\begin{tabular}{|c|c|c|c|c|}\cline{3-5}
			\multicolumn{2}{c|}{}&$N = 50$&$N = 100$&$N = 200$\\ \hline
			{$T$ vs. $U_{X}$}&linear&$3.8$&$3.794$&$3.793$\\ \cline{2-5}
			&quadratic&$3.825$&$3.806$&$3.808$\\ \cline{1-5}
			{$T_{X}$ vs. $U$}&linear&$3.779$&$3.775$&$3.775$\\ \cline{2-5}
			&quadratic&$3.809$&$3.795$&$3.789$\\ \cline{1-5}
		\end{tabular}
	\end{center}
	\caption{Values of the critical interaction strength $U_c$ obtained from the linear and quadratic extrapolation of the crossover plots in Fig.~\ref{fig:crossover} for various values of $N$.}
	\label{table:crossover}
\end{table}

{\em{Critical exponent $z$ and an alternate determination of $U_{c}$}}. We can find the critical value $U_c$ using another approach. This approach lets us estimate the dynamical critical exponent $z$ also. In Fig.~\ref{fig:log_xi_log_T_U} we plot $\ln \xi$ as a function of $\ln T$ for $N = 100$ and $200$ where the correlation length is sufficiently small that finite-size errors are not important (except far from $U_c$). For $U < U_{c}$, $\ln \xi$ saturates at low temperatures, while for $U > U_{c}$, $\ln \xi$ diverges and finally, at $U_{c}$, $\xi$ has a pure power law behavior. In order to determine $U_c$, we fit $\ln \xi$ versus $\ln T$ for various values of $U$ with straight lines. The value of $U$ that gives the best fit is taken as $U_{c}$. It is the slope of $\ln \xi$ vs $\ln T$ that gives us the numerical estimate of the dynamical critical exponent $z$. Despite the fact that we are not in the asymptotic regime for $\xi_{0}$ and $\Gamma_{0}$, the value so obtained is $z = 1.00$ for $U_{c} = 3.8 \pm 0.
005$. For high temperatures, all curves have the same slope as the case $U = U_{c}$. For $N = 50$, where we saw finite-size effects in Fig.~\ref{fig:crossover}, the largest correlation length is close to $N/2$ at the smallest temperature for $U=U_c$, invalidating the estimate. Indeed, in that case $U_{c} = 3.85 \pm 0.005$ but $z = 0.87$, which is clearly incorrect. 

Taking the average value of $U_c$ obtained for $N=100$ and $N=200$ in Table \ref{table:crossover} and estimating the error from the range of values obtained, we find that $U_c=3.79\pm0.01$, consistent with the result obtained from the estimate of the previous paragraph with $z=1$. 

\section{CONCLUSION}\label{sec:6}
The nonperturbative TPSC theory has been extended to a multi-band case, namely the Hubbard model on the honeycomb lattice. In TPSC, valid from weak to intermediate coupling, charge and spin irreducible interactions are determined self-consistently in such a way that conservation laws and the Pauli principle are satisfied. The Mermin-Wagner theorem is also automatically satisfied and the physics of Kanamori-Brueckner screening that renormalizes the spin and charge irreducible vertices is taken into account. On the honeycomb lattice, nearest-neighbor antiferromagnetic fluctuations are dominant. The TPSC value of $U_{c}/t$ for the quantum-critical semimetallic to antiferromagnetic transition is $U_c/t=3.79\pm0.01$ consistent with~\cite{sorella2012} $U_c/t=3.869 \pm 0.013$  and~\cite{pinning_assaad_2013} $U_c/t=3.78$ obtained from the large scale quantum Monte Carlo calculations and also consistent with the functional renormalization group~\cite{honerkamp_density_2008,raghu_topological_2008} $U_c/t=3.8$. These 
results rule out the existence of a spin liquid phase in the ground state of the graphene Hubbard model at intermediate couplings since estimates for the Mott transition yield a $U_{Mott}$ larger than $U_c$. We have also estimated the crossover line in the $T$-$U$ plane where one enters the renormalized classical regime and where a pseudogap is expected to open up. 

Generalized extensions of TPSC to multiband cases of the type presented here and in Ref.~\onlinecite{AritaTPSC:2013} have the potential to open the study of interacting systems, and to improve realistic materials calculations. In the latter case, TPSC offers the possibility to include long wave length spin fluctuations in addition to long wave length charge fluctuations already present in these approaches. 

\acknowledgments
We are grateful to Dominic Bergeron and Wei Wu for illuminating discussions. S.A. would like to thank P. Mangalapandi for his expert advice on parallel computation. This work was supported by the Natural Sciences and Engineering Research Council of Canada (NSERC), and by the Tier I Canada Research Chair Program (A.-M.S.T.).
\vspace{0.5cm}
\appendix

\section{NONINTERACTING SUSCEPTIBILITIES}\label{non-interacting}
Consider the spin susceptibility 
\begin{align}
\chi_{\alpha\beta}(1,2) & = \langle {\mathrm{T}_\tau S^z_{\alpha}(1) S^z_\beta(2)}\rangle
\end{align}
where $\alpha,\beta = a,b$ and $S^z_{\alpha}(1)=n_{\uparrow,\alpha}(1)-n_{\downarrow,\alpha}(1)$. 
Evaluating this expression in terms of noninteracting Green functions in Eq.~\eqref{lf}, we find 
\begin{align}
\chi^0_{aa}({\bf {q}},i\nu) & = -\frac{2}{N^2} \sum_{\bf{k},\alpha,\beta} \frac{1}{4} \left[ M_{\alpha\beta}(\mathbf{k},\mathbf{q}, i\nu)\right]
\end{align}
where $\nu$ is a Matsubara frequency and
\begin{align}
M_{\alpha\beta}({\bf{k}},{\bf{q}},i\nu) & = \frac{n(E_{\bf{k}}^\alpha) - n(E_{\bf{k+q}}^\beta)}{i\nu + E_{\bf{k}}^\alpha -E_{\bf{k+q}}^\beta}
\end{align}
and $\alpha,\beta = \pm$,  with the Fermi function
\begin{align}
n(E^\alpha_{\bf{k}}) & = \frac{1}{e^{E^\alpha_{\bf{k}} /T} + 1}.
\end{align}
and eigenenergies 
\begin{align}
E^\alpha_{\bf{k}} & = \alpha |f(\bf{k})| \\
&= \alpha \sqrt{3+2\cos k_1 +2\cos k_2+2\cos(-k_1-k_2)}. 
\end{align}
$f(\mathbf{k})$ is defined in Eq.~\eqref{f_k}, $T$ is the temperature, while $k_1$ and $k_2$ are the components of the momentum vector on the two unit lattice vectors $\mathbf{a}_1$ and $\mathbf{a}_2$. The other components of the susceptibility tensor are given by
\begin{widetext}
\begin{align}
\chi^0_{ab}({\bf {q}},i\nu) & = -\frac{2}{N^2} \sum_{{\bf{k}},\alpha,\beta} \frac{1}{4} e^{i(\phi_{\bf{k+q}}-\phi_{\bf{k}})} \left[ \alpha\beta M_{\alpha\beta}({\bf{k}},{\bf{q}},i\nu) \right]\label{chi_0_ab}
\end{align}
where $e^{i\phi_{\bf{k}}} = \frac{f(\bf{k})}{|f(\bf{k})|}$ and
\begin{align}
\chi^0_{ba}({\bf {q}},i\nu) & = -\frac{2}{N^2} \sum_{{\bf{k}}} \frac{1}{4} e^{-i(\phi_{\bf{k+q}}-\phi_{\bf{k}})} \left[\alpha\beta M_{\alpha,\beta}({\bf{k}},{\bf{q}},i\nu) \right].
\end{align}
The relation between $\chi^0_{ba}$ and $\chi^0_{ab}$  in Eq.~\eqref{chi_ab-Identity} is thus satisfied. 
\end{widetext}
The equivalence of the two sublattices implies $\chi^0_{aa} = \chi^0_{bb}$ but in general not $\chi^0_{ab} = \chi^0_{ba}$ because of the chiral nature of the Dirac points. 

Instead of using FFT's,  one can first perform the Matsubara frequency sum exactly and then sum over wave vectors. In that case, there are certain points in the Brillouin zone where the $M_{\alpha\beta}$ have the form $0/0$. At these points one must take the limits and use l'Hospital's rule. For example at ${\bf q} = 0$
\begin{align}
\lim_{{\bf q}\rightarrow {\bf 0}} M_{++}({\bf{k}},{\bf{q}},0) & = \lim_{{\bf q}\rightarrow {\bf 0}}  \frac{n(E_k^+) - n(E_{\bf{k+q}}^+)}{E_k^+ -E_{\bf{k+q}}^+} \\
& = \lim_{{\bf q}\rightarrow {\bf 0}}  \frac{\partial  n(E_{\bf{k+q}}^+)}{\partial E_{\bf{k+q}}^+} \\
& = -\beta n(E_{\bf{k}}^+) \left[1-n(E_{\bf{k}}^+)\right].
\end{align}
Similarly $M_{\alpha\beta} = 0/0$ at $\mathbf{k} = (2\pi/3,2\pi/3)$ and $\mathbf{q} = (2\pi/3,2\pi/3)$ so the same solution applies. However, this procedure means that the Dirac points introduce large finite-size effects in the temperature dependence. Choosing a grid that is regular but avoids the Dirac points (for example $N=100$ instead of $N=90$) minimizes finite-size effects.  

\section{ANTIFERROMAGNETIC SUSCEPTIBILITY}\label{appendix_AFM}
In the presence of interactions the spin susceptibility is given by the matrix equation
\begin{align}
\boldsymbol{\chi}^{s} & =   \left(\mathbf{I} - \frac{\mathcal{U}_s}{2}\boldsymbol{\chi}^0 \right)^{-1} \boldsymbol{\chi}^0.
\end{align}
Defining the antiferromagnetic susceptibility by
\begin{align}
\chi_{afm}({\bf {q}},i\nu) & = \chi_{aa}({\bf {q}},i\nu) + \sqrt{\chi_{ab}({\bf {q}},i\nu)\chi_{ba}({\bf {q}},i\nu)}
\end{align}
allows us to find a simple scalar equation that reduces to $\chi_{afm} = \chi_{aa} + \chi_{ab}$ when $\chi_{ab}$ is real. The combination $ \chi_{aa} + \chi_{ab}$ does not satisfy a simple scalar equation in the general case. 

The algebra that follows proves these assertions. Expanding the matrix equation we find
\begin{widetext}
\begin{align}
\boldsymbol{\chi}^{s} & = \frac{1}{\det} 
\left(\begin{array}{c|c}
1 - \frac{U_s}{2}\chi_{bb}^0 & \frac{U_s}{2}\chi_{ab}^0 \\
\hline
\frac{U_s}{2} \chi_{ba}^0 & 1 - \frac{U_s}{2}\chi_{aa}^0 
\end{array}\right) \left(\begin{array}{c|c}
\chi_{aa}^0 & \chi_{ab}^0 \\
\hline
\chi_{ba}^0 & \chi_{bb}^0 
\end{array}\right)\\
& = \frac{1}{\det} 
\left(\begin{array}{c|c}
(1 - \frac{U_s}{2}\chi_{bb}^0)\chi_{aa}^0 + \frac{U_s}{2}\chi_{ab}^0\chi_{ba}^0 & \chi_{ab}^0 \\
\hline
\chi_{ba}^0 & \frac{U_s}{2} \chi_{ba}^0 \chi_{ab}^0 + (1 - \frac{U_s}{2}\chi_{aa}^0 )\chi_{bb}^0
\end{array}\right) 
\end{align}
where  $\det$, that stands for the determinant, can be expanded as
\begin{align}
\det &= (1 - \frac{U_s}{2}\chi_{bb}^0)(1 - \frac{U_s}{2}\chi_{aa}^0 ) - \frac{U_s^2}{4}\chi_{ab}^0\chi_{ba}^0 \\
& = 1 - \frac{U_s}{2}(\chi_{bb}^0 + \chi_{aa}^0) - \frac{U_s^2}{4}(\chi_{ab}^0\chi_{ba}^0  - \chi^0_{aa}\chi^0_{bb}) \\
& = \left(1 - \frac{U_s}{2} (\sqrt{\chi_{aa}^0\chi_{bb}^0} + \sqrt{\chi_{ab}^0\chi_{ba}^0})\right)\left(1 - \frac{U_s}{2} (\sqrt{\chi_{aa}^0\chi_{bb}^0} - \sqrt{\chi_{ab}^0\chi_{ba}^0})\right).\\
\end{align}
With $\chi^0_{aa} = \chi^0_{bb}$ we can simplify the determinant
\begin{align}
\det & = \left(1 - \frac{U_s}{2} (\chi_{aa}^0 + \sqrt{\chi_{ab}^0\chi_{ba}^0})\right)\left(1 - \frac{U_s}{2} (\chi_{aa}^0 - \sqrt{\chi_{ab}^0\chi_{ba}^0})\right)
\end{align}
and the antiferromagnetic susceptibility
\begin{align}
\chi^{s}_{afm} & = \chi^{s}_{aa} + \sqrt{\chi^{s}_{ab}\chi^{s}_{ba}}\\
& = \frac{(1 - \frac{U_s}{2}\chi_{bb}^0)\chi_{aa}^0 + \frac{U_s}{2}\chi_{ab}^0\chi_{ba}^0 + \sqrt{\chi_{ab}^0\chi_{ba}^0}}{\left(1 - \frac{U_s}{2} (\chi_{aa}^0 + \sqrt{\chi_{ab}^0\chi_{ba}^0})\right)\left(1 - \frac{U_s}{2} (\chi_{aa}^0 - \sqrt{\chi_{ab}^0\chi_{ba}^0})\right)} \\
& = \frac{\chi_{aa}^0  + \sqrt{\chi_{ab}^0\chi_{ba}^0}- \frac{U_s}{2}((\chi_{aa}^{0})^2 - \chi_{ab}^0\chi_{ba}^0)}{\left(1 - \frac{U_s}{2} (\chi_{aa}^0 +\sqrt{\chi_{ab}^0\chi_{ba}^0})\right)\left(1 - \frac{U_s}{2} (\chi_{aa}^0 - \sqrt{\chi_{ab}^0\chi_{ba}^0})\right)} \\
& = \frac{\left(\chi_{aa}^0  + \sqrt{\chi_{ab}^0\chi_{ba}^0}\right)(1 - \frac{U_s}{2}(\chi_{aa}^0 - \sqrt{\chi_{ab}^0\chi_{ba}^0}))}{\left(1 - \frac{U_s}{2} (\chi_{aa}^0 +\sqrt{\chi_{ab}^0\chi_{ba}^0})\right)\left(1 - \frac{U_s}{2} (\chi_{aa}^0 - \sqrt{\chi_{ab}^0\chi_{ba}^0})\right)} \\
& = \frac{\chi_{aa}^0  + \sqrt{\chi_{ab}^0\chi_{ba}^0}}{\left(1 - \frac{U_s}{2} (\chi_{aa}^0 +\sqrt{\chi_{ab}^0\chi_{ba}^0})\right)} 
\end{align}
\end{widetext}
so that
\begin{align}
\chi^{s}_{afm}({\bf {q}},i\nu) & = \frac{\chi^0_{afm}({\bf {q}},i\nu) }{1 - \frac{U_s}{2} \chi^0_{afm}({\bf {q}},i\nu)}
\end{align}
\section{DERIVATIVES OF NONINTERACTING SUSCEPTIBILITIES IN THE DIRAC APPROXIMATION AND ESTIMATES FOR $\xi_0,\Gamma_0$ }\label{Appendix_analytic_dirac}

\subsection{$\chi^0_{afm}$ and its derivatives}
\begin{widetext}
At $\mathbf{q}=0$, the phase in $\chi^0_{ab}$, Eq.~\eqref{chi_0_ab}, disappears and we are left with $\chi^0_{ab}(\mathbf{q}=0,i\nu)=\chi^0_{ba}(\mathbf{q}=0,i\nu)$.  Given that we are taking the positive square root, we also have $\sqrt{\chi_{ab}(0,i\nu)\chi_{ba}(0,i\nu)}=-\chi_{ab}(0,i\nu)$ so that the retarded function is

\begin{align}
\chi^0_{afm}(0,\omega+i\delta) & =\chi^0_{aa}(0,\omega+i\delta) -\chi^0_{ab}(0,\omega+i\delta) \\
		&= -\frac{1}{N^2} \sum_{\bf{k}} \left[ M_{+-}(\mathbf{k},\mathbf{q}, \omega+i\delta)+M_{-+}(\mathbf{k},\mathbf{q}, \omega+i\delta)\right]\\
		&= -\frac{1}{N^2} \sum_{\bf{k}} \left[ \frac{2n(E^+_{\mathbf{k}})-1}{2E^+_{\mathbf{k}}+\omega+i\delta}+\frac{-2n(E^+_{\mathbf{k}})+1}{-2E^+_{\mathbf{k}}+\omega+i\delta}\right].
\end{align}
Only interband transitions contribute to $\chi^0_{afm}(0,i\nu)$. 

In the Dirac approximation, we evaluate separately the real and imaginary parts. Beginning with the latter, we find
\begin{equation}
\operatorname{Im}\chi_{afm}^{0}\left(  0,\omega+i\delta\right)  =-\pi\frac
{1}{N^{2}}\sum_{\mathbf{k}}\tanh\left(  \beta E^+_{\mathbf{k}}/2\right)  \left(  \delta\left(  2E^+_{\mathbf{k}}+\omega\right)  -\delta\left(  2E^+_{\mathbf{k}}-\omega\right)  \right).
\end{equation}
Transforming the sum into an integral, going to cylindrical coordinates, we
have%
\begin{equation}
\frac{1}{N^{2}}\sum_{\mathbf{k}}\rightarrow\int\frac{d^{2}k}{\left(
	2\pi\right)  ^{2}}\rightarrow\int_{0}^{\Lambda}\frac{kdk}{2\pi}\rightarrow
\frac{1}{v_{F}^{2}}\int_{0}^{\Lambda_{E}}\frac{\varepsilon d\varepsilon}{2\pi
}
\end{equation}
where $\Lambda_E$ is the energy cutoff. Assuming $\omega>0,$ only the last $\delta$ function contributes. Taking into
account a factor of $2$ for the two Dirac points, we are left with%
\begin{equation}
\operatorname{Im}\chi_{afm}^{0}\left(  0,\omega+i\delta\right)  =\pi\frac
{2}{v_{F}^{2}}\int_{0}^{\Lambda_{E}}\frac{\varepsilon d\varepsilon}{2\pi}%
\tanh\left(  \beta\varepsilon/2\right)  \delta\left(  2\varepsilon
-\omega\right)  .
\end{equation}
In the zero temperature limit,%
\begin{align}
\operatorname{Im}\chi_{afm}^{0}\left(  0,\omega+i\delta\right)   &  =\frac
{1}{v_{F}^{2}}\int_{0}^{\Lambda_{E}}\delta\left(  2\varepsilon-\omega\right)
\varepsilon d\varepsilon\\
&  =\frac{1}{2v_{F}^{2}}\frac{\omega}{2}.%
\label{Im_chi0}
\end{align}

For the real part, we begin with
\begin{equation}
\operatorname{Re}\chi_{afm}^{0}\left(  0,\omega+i\delta\right)  =\frac
{1}{N^{2}}\mathcal{P}\sum_{\mathbf{k}}\tanh\left(  \beta
E^+_{\mathbf{k}}/2\right)  \frac{4E^+_{\mathbf
		{k}}}{\left(  2E^+_{\mathbf{k}}\right)  ^{2}-\omega^{2}}.%
\end{equation}
Expanding around the two Dirac points as above gives 
\begin{equation}
\operatorname{Re}\chi_{afm}^{0}\left(  0,\omega+i\delta\right)  =\mathcal{P}%
\frac{2}{v_{F}^{2}}\int_{0}^{\Lambda_{E}}\frac{\varepsilon d\varepsilon}{2\pi
}\tanh\left(  \beta\varepsilon/2\right)  \frac{4\varepsilon}{\left(
2\varepsilon\right)  ^{2}-\omega^{2}}.
\end{equation}
Working in the zero temperature limit, we find%

\begin{align}
\operatorname{Re}\chi_{afm}^{0}\left(  0,\omega+i\delta\right)   &
=\mathcal{P}\frac{1}{v_{F}^{2}}\int_{0}^{\Lambda_{E}}\frac{d\varepsilon}{\pi
}\frac{\left(  2\varepsilon\right)  ^{2}}{\left(  2\varepsilon\right)
^{2}-\omega^{2}}\\
&  =\frac{1}{v_{F}^{2}}\int_{0}^{\Lambda_{E}}\frac{d\varepsilon}{\pi}%
+\omega^{2}\frac{1}{v_{F}^{2}}\mathcal{P}\int_{0}^{\Lambda_{E}}\frac
{d\varepsilon}{\pi}\frac{1}{\left(  2\varepsilon\right)  ^{2}-\omega^{2}}\\
&  =\frac{1}{\pi v_{F}^{2}}\Lambda_{E}+\frac{\omega^{2}}{2\omega v_{F}^{2}\pi
}\mathcal{P}\int_{0}^{2\Lambda_{E}/\omega}dx\frac{1}{x^{2}-1}.%
\end{align}
Assuming $\omega>0,$ we find%
\begin{align}
\operatorname{Re}\chi_{afm}^{0}\left(  0,\omega+i\delta\right)   &  =\frac
{1}{\pi v_{F}^{2}}\Lambda_{E}-\frac{\omega}{2v_{F}^{2}\pi}\tanh^{-1}\left(
\frac{\omega}{2\Lambda_{E}}\right)  \\
&  \sim\frac{1}{\pi v_{F}^{2}}\Lambda_{E}-\frac{\omega^{2}}{4v_{F}^{2}%
	\pi\Lambda_{E}}\label{Re_chi0}.%
\end{align}

\subsection{Estimates for $\xi_0$ and $\Gamma_0$}
We begin with the definition Eq.~\eqref{Def_Gamma0} of $\Gamma_0$ and substitute the results just found Eqs.~\eqref{Im_chi0} and \eqref{Re_chi0} to find
\begin{align}
\frac{1}{\Gamma_{0}} &  =\frac{1}{\xi_{0}}\frac{1}{\operatorname{Re}\chi
	_{afm}^{0}\left(  0,0\right)  }\frac{\partial}{\partial\omega}%
	\operatorname{Im}\chi_{afm}^{0}\left(  0,\omega+i\delta\right)  \\
	&  =\frac{1}{\xi_{0}}\frac{\frac{1}{4v_{F}^{2}}}{\frac{1}{\pi v_{F}^{2}%
			}\Lambda_{E}}=\frac{\pi}{4\xi_{0}\Lambda_{E}}%
			\end{align}
\end{widetext}
so as expected $\Gamma_{0}$ has units of velocity since with $\hbar=1,$ $\Lambda_{E}$ is $\left(  time\right)  ^{-1}.$ Taking $\Lambda_{E}=v_{F}\Lambda$ with $\Lambda=\pi/a$ the cutoff, then%
			\begin{equation}
			\Gamma_{0}=\frac{4\xi_{0}\Lambda_{E}}{\pi}=4\frac{\xi_{0}}{a}v_{F}.
			\end{equation}
From the Lorentz invariance, we expect 
			\begin{equation}
			\Gamma_{0}= v_{F}%
			\end{equation}
which, with $a=1$, suggests that $\xi_{0}\sim 0.25$. The result found numerically for $\xi_{0}$ in Fig.~\ref{fig:xi0_G0inv} is just slightly larger because band curvature means that $\Lambda_{E}$ is a bit smaller than the estimate $\Lambda_{E}=v_{F}\Lambda$. Similarly, $\Gamma_{0}$ at low temperatures is numerically close to $v_{F}=\sqrt{3}/2$ in our units.


\begin{thebibliography}{58}%
	\makeatletter
	\providecommand \@ifxundefined [1]{%
		\@ifx{#1\undefined}
	}%
	\providecommand \@ifnum [1]{%
		\ifnum #1\expandafter \@firstoftwo
		\else \expandafter \@secondoftwo
		\fi
	}%
	\providecommand \@ifx [1]{%
		\ifx #1\expandafter \@firstoftwo
		\else \expandafter \@secondoftwo
		\fi
	}%
	\providecommand \natexlab [1]{#1}%
	\providecommand \enquote  [1]{``#1''}%
	\providecommand \bibnamefont  [1]{#1}%
	\providecommand \bibfnamefont [1]{#1}%
	\providecommand \citenamefont [1]{#1}%
	\providecommand \href@noop [0]{\@secondoftwo}%
	\providecommand \href [0]{\begingroup \@sanitize@url \@href}%
	\providecommand \@href[1]{\@@startlink{#1}\@@href}%
	\providecommand \@@href[1]{\endgroup#1\@@endlink}%
	\providecommand \@sanitize@url [0]{\catcode `\\12\catcode `\$12\catcode
		`\&12\catcode `\#12\catcode `\^12\catcode `\_12\catcode `\%12\relax}%
	\providecommand \@@startlink[1]{}%
	\providecommand \@@endlink[0]{}%
	\providecommand \url  [0]{\begingroup\@sanitize@url \@url }%
	\providecommand \@url [1]{\endgroup\@href {#1}{\urlprefix }}%
	\providecommand \urlprefix  [0]{URL }%
	\providecommand \Eprint [0]{\href }%
	\providecommand \doibase [0]{http://dx.doi.org/}%
	\providecommand \selectlanguage [0]{\@gobble}%
	\providecommand \bibinfo  [0]{\@secondoftwo}%
	\providecommand \bibfield  [0]{\@secondoftwo}%
	\providecommand \translation [1]{[#1]}%
	\providecommand \BibitemOpen [0]{}%
	\providecommand \bibitemStop [0]{}%
	\providecommand \bibitemNoStop [0]{.\EOS\space}%
	\providecommand \EOS [0]{\spacefactor3000\relax}%
	\providecommand \BibitemShut  [1]{\csname bibitem#1\endcsname}%
	\let\auto@bib@innerbib\@empty
	\bibitem [{\citenamefont {Anderson}(1987)}]{Anderson:1987}%
	\BibitemOpen
	\bibfield  {author} {\bibinfo {author} {\bibfnamefont {P.~W.}\ \bibnamefont
			{Anderson}},\ }\href
	{http://www.sciencemag.org/content/235/4793/1196.abstract} {\bibfield
		{journal} {\bibinfo  {journal} {Science}\ }\textbf {\bibinfo {volume}
			{235}},\ \bibinfo {pages} {1196 } (\bibinfo {year} {1987})}\BibitemShut
	{NoStop}%
	\bibitem [{\citenamefont {{Gingras}}\ and\ \citenamefont
		{{McClarty}}(2013)}]{GingrasReview:2013}%
	\BibitemOpen
	\bibfield  {author} {\bibinfo {author} {\bibfnamefont {M.~J.~P.}\
			\bibnamefont {{Gingras}}}\ and\ \bibinfo {author} {\bibfnamefont {P.~A.}\
			\bibnamefont {{McClarty}}},\ }\href@noop {} {\bibfield  {journal} {\bibinfo
			{journal} {ArXiv e-prints}\ } (\bibinfo {year} {2013})},\ \Eprint
	{http://arxiv.org/abs/1311.1817} {arXiv:1311.1817 [cond-mat.str-el]}
	\BibitemShut {NoStop}%
	\bibitem [{\citenamefont {Powell}\ and\ \citenamefont
		{McKenzie}(2011)}]{PowellMcKenzieReview:2011}%
	\BibitemOpen
	\bibfield  {author} {\bibinfo {author} {\bibfnamefont {B.~J.}\ \bibnamefont
			{Powell}}\ and\ \bibinfo {author} {\bibfnamefont {R.~H.}\ \bibnamefont
			{McKenzie}},\ }\href {http://stacks.iop.org/0034-4885/74/i=5/a=056501}
	{\bibfield  {journal} {\bibinfo  {journal} {Reports on Progress in Physics}\
		}\textbf {\bibinfo {volume} {74}},\ \bibinfo {pages} {056501} (\bibinfo
		{year} {2011})}\BibitemShut {NoStop}%
	\bibitem [{\citenamefont {Anderson}(1973)}]{Anderson:1973}%
	\BibitemOpen
	\bibfield  {author} {\bibinfo {author} {\bibfnamefont {P.~W.}\ \bibnamefont
			{Anderson}},\ }\href@noop {} {\bibfield  {journal} {\bibinfo  {journal} {Mat.
				Res. Bull.}\ }\textbf {\bibinfo {volume} {8}},\ \bibinfo {pages} {153}
		(\bibinfo {year} {1973})}\BibitemShut {NoStop}%
	\bibitem [{\citenamefont {Yan}\ \emph {et~al.}(2011)\citenamefont {Yan},
		\citenamefont {Huse},\ and\ \citenamefont {White}}]{yan_spin-liquid_2011}%
	\BibitemOpen
	\bibfield  {author} {\bibinfo {author} {\bibfnamefont {S.}~\bibnamefont
			{Yan}}, \bibinfo {author} {\bibfnamefont {D.~A.}\ \bibnamefont {Huse}}, \
		and\ \bibinfo {author} {\bibfnamefont {S.~R.}\ \bibnamefont {White}},\ }\href
	{\doibase 10.1126/science.1201080} {\bibfield  {journal} {\bibinfo  {journal}
			{Science}\ }\textbf {\bibinfo {volume} {332}},\ \bibinfo {pages} {1173}
		(\bibinfo {year} {2011})}\BibitemShut {NoStop}%
	\bibitem [{\citenamefont {Paiva}\ \emph {et~al.}(2005)\citenamefont {Paiva},
		\citenamefont {Scalettar}, \citenamefont {Zheng}, \citenamefont {Singh},\
		and\ \citenamefont {Oitmaa}}]{paiva2005}%
	\BibitemOpen
	\bibfield  {author} {\bibinfo {author} {\bibfnamefont {T.}~\bibnamefont
			{Paiva}}, \bibinfo {author} {\bibfnamefont {R.~T.}\ \bibnamefont
			{Scalettar}}, \bibinfo {author} {\bibfnamefont {W.}~\bibnamefont {Zheng}},
		\bibinfo {author} {\bibfnamefont {R.~R.~P.}\ \bibnamefont {Singh}}, \ and\
		\bibinfo {author} {\bibfnamefont {J.}~\bibnamefont {Oitmaa}},\ }\href
	{\doibase 10.1103/PhysRevB.72.085123} {\bibfield  {journal} {\bibinfo
			{journal} {Phys. Rev. B}\ }\textbf {\bibinfo {volume} {72}},\ \bibinfo
		{pages} {085123} (\bibinfo {year} {2005})}\BibitemShut {NoStop}%
	\bibitem [{\citenamefont {Meng}\ \emph {et~al.}(2010)\citenamefont {Meng},
		\citenamefont {Lang}, \citenamefont {Wessel}, \citenamefont {Assaad},\ and\
		\citenamefont {Muramatsu}}]{meng2010}%
	\BibitemOpen
	\bibfield  {author} {\bibinfo {author} {\bibfnamefont {Z.}~\bibnamefont
			{Meng}}, \bibinfo {author} {\bibfnamefont {T.}~\bibnamefont {Lang}}, \bibinfo
		{author} {\bibfnamefont {S.}~\bibnamefont {Wessel}}, \bibinfo {author}
		{\bibfnamefont {F.}~\bibnamefont {Assaad}}, \ and\ \bibinfo {author}
		{\bibfnamefont {A.}~\bibnamefont {Muramatsu}},\ }\href
	{http://dx.doi.org/10.1038/nature08942} {\bibfield  {journal} {\bibinfo
			{journal} {Nature}\ }\textbf {\bibinfo {volume} {464}},\ \bibinfo {pages}
		{847} (\bibinfo {year} {2010})}\BibitemShut {NoStop}%
	\bibitem [{\citenamefont {Hohenadler}\ \emph {et~al.}(2011)\citenamefont
		{Hohenadler}, \citenamefont {Lang},\ and\ \citenamefont
		{Assaad}}]{hohenadler_correlation_2011}%
	\BibitemOpen
	\bibfield  {author} {\bibinfo {author} {\bibfnamefont {M.}~\bibnamefont
			{Hohenadler}}, \bibinfo {author} {\bibfnamefont {T.~C.}\ \bibnamefont
			{Lang}}, \ and\ \bibinfo {author} {\bibfnamefont {F.~F.}\ \bibnamefont
			{Assaad}},\ }\href {\doibase 10.1103/PhysRevLett.106.100403} {\bibfield
		{journal} {\bibinfo  {journal} {Physical Review Letters}\ }\textbf {\bibinfo
			{volume} {106}},\ \bibinfo {pages} {100403} (\bibinfo {year}
		{2011})}\BibitemShut {NoStop}%
	\bibitem [{\citenamefont {Hohenadler}\ \emph
		{et~al.}(2012{\natexlab{a}})\citenamefont {Hohenadler}, \citenamefont
		{Lang},\ and\ \citenamefont {Assaad}}]{hohenadler_erratum:_2012}%
	\BibitemOpen
	\bibfield  {author} {\bibinfo {author} {\bibfnamefont {M.}~\bibnamefont
			{Hohenadler}}, \bibinfo {author} {\bibfnamefont {T.~C.}\ \bibnamefont
			{Lang}}, \ and\ \bibinfo {author} {\bibfnamefont {F.~F.}\ \bibnamefont
			{Assaad}},\ }\href {\doibase 10.1103/PhysRevLett.109.229902} {\bibfield
		{journal} {\bibinfo  {journal} {Physical Review Letters}\ }\textbf {\bibinfo
			{volume} {109}},\ \bibinfo {pages} {229902} (\bibinfo {year}
		{2012}{\natexlab{a}})}\BibitemShut {NoStop}%
	\bibitem [{\citenamefont {Hohenadler}\ \emph
		{et~al.}(2012{\natexlab{b}})\citenamefont {Hohenadler}, \citenamefont {Meng},
		\citenamefont {Lang}, \citenamefont {Wessel}, \citenamefont {Muramatsu},\
		and\ \citenamefont {Assaad}}]{hohenadler_quantum_2012}%
	\BibitemOpen
	\bibfield  {author} {\bibinfo {author} {\bibfnamefont {M.}~\bibnamefont
			{Hohenadler}}, \bibinfo {author} {\bibfnamefont {Z.~Y.}\ \bibnamefont
			{Meng}}, \bibinfo {author} {\bibfnamefont {T.~C.}\ \bibnamefont {Lang}},
		\bibinfo {author} {\bibfnamefont {S.}~\bibnamefont {Wessel}}, \bibinfo
		{author} {\bibfnamefont {A.}~\bibnamefont {Muramatsu}}, \ and\ \bibinfo
		{author} {\bibfnamefont {F.~F.}\ \bibnamefont {Assaad}},\ }\href {\doibase
		10.1103/PhysRevB.85.115132} {\bibfield  {journal} {\bibinfo  {journal}
			{Physical Review B}\ }\textbf {\bibinfo {volume} {85}},\ \bibinfo {pages}
		{115132} (\bibinfo {year} {2012}{\natexlab{b}})}\BibitemShut {NoStop}%
	\bibitem [{\citenamefont {Zheng}\ \emph {et~al.}(2011)\citenamefont {Zheng},
		\citenamefont {Zhang},\ and\ \citenamefont {Wu}}]{zheng_particle-hole_2011}%
	\BibitemOpen
	\bibfield  {author} {\bibinfo {author} {\bibfnamefont {D.}~\bibnamefont
			{Zheng}}, \bibinfo {author} {\bibfnamefont {G.-M.}\ \bibnamefont {Zhang}}, \
		and\ \bibinfo {author} {\bibfnamefont {C.}~\bibnamefont {Wu}},\ }\href
	{\doibase 10.1103/PhysRevB.84.205121} {\bibfield  {journal} {\bibinfo
			{journal} {Physical Review B}\ }\textbf {\bibinfo {volume} {84}},\ \bibinfo
		{pages} {205121} (\bibinfo {year} {2011})}\BibitemShut {NoStop}%
	\bibitem [{\citenamefont {Sorella}\ \emph {et~al.}(2012)\citenamefont
		{Sorella}, \citenamefont {Otsuka},\ and\ \citenamefont
		{Yunoki}}]{sorella2012}%
	\BibitemOpen
	\bibfield  {author} {\bibinfo {author} {\bibfnamefont {S.}~\bibnamefont
			{Sorella}}, \bibinfo {author} {\bibfnamefont {Y.}~\bibnamefont {Otsuka}}, \
		and\ \bibinfo {author} {\bibfnamefont {S.}~\bibnamefont {Yunoki}},\ }\href
	{http://dx.doi.org/10.1038/srep00992} {\bibfield  {journal} {\bibinfo
			{journal} {Scientific reports}\ }\textbf {\bibinfo {volume} {2}} (\bibinfo
		{year} {2012})}\BibitemShut {NoStop}%
	\bibitem [{\citenamefont {Maier}\ \emph {et~al.}(2005)\citenamefont {Maier},
		\citenamefont {Jarrell}, \citenamefont {Pruschke},\ and\ \citenamefont
		{Hettler}}]{Maier:2005}%
	\BibitemOpen
	\bibfield  {author} {\bibinfo {author} {\bibfnamefont {T.}~\bibnamefont
			{Maier}}, \bibinfo {author} {\bibfnamefont {M.}~\bibnamefont {Jarrell}},
		\bibinfo {author} {\bibfnamefont {T.}~\bibnamefont {Pruschke}}, \ and\
		\bibinfo {author} {\bibfnamefont {M.~H.}\ \bibnamefont {Hettler}},\ }\href
	{\doibase 10.1103/RevModPhys.77.1027} {\bibfield  {journal} {\bibinfo
			{journal} {Rev. Mod. Phys.}\ }\textbf {\bibinfo {volume} {77}},\ \bibinfo
		{pages} {1027} (\bibinfo {year} {2005})}\BibitemShut {NoStop}%
	\bibitem [{\citenamefont {Kotliar}\ \emph {et~al.}(2006)\citenamefont
		{Kotliar}, \citenamefont {Savrasov}, \citenamefont {Haule}, \citenamefont
		{Oudovenko}, \citenamefont {Parcollet},\ and\ \citenamefont
		{Marianetti}}]{KotliarRMP:2006}%
	\BibitemOpen
	\bibfield  {author} {\bibinfo {author} {\bibfnamefont {G.}~\bibnamefont
			{Kotliar}}, \bibinfo {author} {\bibfnamefont {S.~Y.}\ \bibnamefont
			{Savrasov}}, \bibinfo {author} {\bibfnamefont {K.}~\bibnamefont {Haule}},
		\bibinfo {author} {\bibfnamefont {V.~S.}\ \bibnamefont {Oudovenko}}, \bibinfo
		{author} {\bibfnamefont {O.}~\bibnamefont {Parcollet}}, \ and\ \bibinfo
		{author} {\bibfnamefont {C.~A.}\ \bibnamefont {Marianetti}},\ }\href
	{\doibase 10.1103/RevModPhys.78.865} {\bibfield  {journal} {\bibinfo
			{journal} {Reviews of Modern Physics}\ }\textbf {\bibinfo {volume} {78}},\
		\bibinfo {eid} {865} (\bibinfo {year} {2006})}\BibitemShut {NoStop}%
	\bibitem [{\citenamefont {Tremblay}\ \emph {et~al.}(2006)\citenamefont
		{Tremblay}, \citenamefont {Kyung},\ and\ \citenamefont
		{S\'en\'echal}}]{LTP:2006}%
	\BibitemOpen
	\bibfield  {author} {\bibinfo {author} {\bibfnamefont {A.~M.~S.}\
			\bibnamefont {Tremblay}}, \bibinfo {author} {\bibfnamefont {B.}~\bibnamefont
			{Kyung}}, \ and\ \bibinfo {author} {\bibfnamefont {D.}~\bibnamefont
			{S\'en\'echal}},\ }\href {http://dx.doi.org/10.1063/1.2199446} {\bibfield
		{journal} {\bibinfo  {journal} {Low Temp. Phys.}\ }\textbf {\bibinfo {volume}
			{32}},\ \bibinfo {pages} {424} (\bibinfo {year} {2006})}\BibitemShut
	{NoStop}%
	\bibitem [{\citenamefont {Tran}\ and\ \citenamefont
		{Kuroki}(2009)}]{tran_finite-temperature_2009}%
	\BibitemOpen
	\bibfield  {author} {\bibinfo {author} {\bibfnamefont {M.-T.}\ \bibnamefont
			{Tran}}\ and\ \bibinfo {author} {\bibfnamefont {K.}~\bibnamefont {Kuroki}},\
	}\href {\doibase 10.1103/PhysRevB.79.125125} {\bibfield  {journal} {\bibinfo
		{journal} {Physical Review B}\ }\textbf {\bibinfo {volume} {79}},\ \bibinfo
	{pages} {125125} (\bibinfo {year} {2009})}\BibitemShut {NoStop}%
\bibitem [{\citenamefont {Jafari}(2009)}]{jafari_dynamical_2009}%
\BibitemOpen
\bibfield  {author} {\bibinfo {author} {\bibfnamefont {S.~A.}\ \bibnamefont
		{Jafari}},\ }\href {\doibase 10.1140/epjb/e2009-00128-1} {\bibfield
	{journal} {\bibinfo  {journal} {The European Physical Journal B}\ }\textbf
	{\bibinfo {volume} {68}},\ \bibinfo {pages} {537} (\bibinfo {year}
	{2009})}\BibitemShut {NoStop}%
\bibitem [{\citenamefont {Ebrahimkhas}(2011)}]{ebrahimkhas_exact_2011}%
\BibitemOpen
\bibfield  {author} {\bibinfo {author} {\bibfnamefont {M.}~\bibnamefont
		{Ebrahimkhas}},\ }\href {\doibase 10.1016/j.physleta.2011.07.019} {\bibfield
	{journal} {\bibinfo  {journal} {Physics Letters A}\ }\textbf {\bibinfo
		{volume} {375}},\ \bibinfo {pages} {3223} (\bibinfo {year}
	{2011})}\BibitemShut {NoStop}%
\bibitem [{\citenamefont {Wu}\ \emph {et~al.}(2012)\citenamefont {Wu},
	\citenamefont {Rachel}, \citenamefont {Liu},\ and\ \citenamefont
	{Le~Hur}}]{wu_quantum_2012}%
\BibitemOpen
\bibfield  {author} {\bibinfo {author} {\bibfnamefont {W.}~\bibnamefont
		{Wu}}, \bibinfo {author} {\bibfnamefont {S.}~\bibnamefont {Rachel}}, \bibinfo
	{author} {\bibfnamefont {W.-M.}\ \bibnamefont {Liu}}, \ and\ \bibinfo
	{author} {\bibfnamefont {K.}~\bibnamefont {Le~Hur}},\ }\href {\doibase
	10.1103/PhysRevB.85.205102} {\bibfield  {journal} {\bibinfo  {journal}
		{Physical Review B}\ }\textbf {\bibinfo {volume} {85}},\ \bibinfo {pages}
	{205102} (\bibinfo {year} {2012})}\BibitemShut {NoStop}%
\bibitem [{\citenamefont {Budich}\ \emph {et~al.}(2013)\citenamefont {Budich},
	\citenamefont {Trauzettel},\ and\ \citenamefont
	{Sangiovanni}}]{budich_fluctuation-driven_2013}%
\BibitemOpen
\bibfield  {author} {\bibinfo {author} {\bibfnamefont {J.~C.}\ \bibnamefont
		{Budich}}, \bibinfo {author} {\bibfnamefont {B.}~\bibnamefont {Trauzettel}},
	\ and\ \bibinfo {author} {\bibfnamefont {G.}~\bibnamefont {Sangiovanni}},\
}\href {\doibase 10.1103/PhysRevB.87.235104} {\bibfield  {journal} {\bibinfo
	{journal} {Physical Review B}\ }\textbf {\bibinfo {volume} {87}},\ \bibinfo
{pages} {235104} (\bibinfo {year} {2013})}\BibitemShut {NoStop}%
\bibitem [{\citenamefont {Yu}\ \emph {et~al.}(2011)\citenamefont {Yu},
	\citenamefont {Xie},\ and\ \citenamefont {Li}}]{yu_mott_2011}%
\BibitemOpen
\bibfield  {author} {\bibinfo {author} {\bibfnamefont {S.-L.}\ \bibnamefont
		{Yu}}, \bibinfo {author} {\bibfnamefont {X.~C.}\ \bibnamefont {Xie}}, \ and\
	\bibinfo {author} {\bibfnamefont {J.-X.}\ \bibnamefont {Li}},\ }\href
{\doibase 10.1103/PhysRevLett.107.010401} {\bibfield  {journal} {\bibinfo
		{journal} {Physical Review Letters}\ }\textbf {\bibinfo {volume} {107}},\
	\bibinfo {pages} {010401} (\bibinfo {year} {2011})}\BibitemShut {NoStop}%
\bibitem [{\citenamefont {Wu}\ \emph {et~al.}(2010)\citenamefont {Wu},
	\citenamefont {Chen}, \citenamefont {Tao}, \citenamefont {Tong},\ and\
	\citenamefont {Liu}}]{WuDirac:2010}%
\BibitemOpen
\bibfield  {author} {\bibinfo {author} {\bibfnamefont {W.}~\bibnamefont
		{Wu}}, \bibinfo {author} {\bibfnamefont {Y.-H.}\ \bibnamefont {Chen}},
	\bibinfo {author} {\bibfnamefont {H.-S.}\ \bibnamefont {Tao}}, \bibinfo
	{author} {\bibfnamefont {N.-H.}\ \bibnamefont {Tong}}, \ and\ \bibinfo
	{author} {\bibfnamefont {W.-M.}\ \bibnamefont {Liu}},\ }\href {\doibase
	10.1103/PhysRevB.82.245102} {\bibfield  {journal} {\bibinfo  {journal} {Phys.
			Rev. B}\ }\textbf {\bibinfo {volume} {82}},\ \bibinfo {pages} {245102}
	(\bibinfo {year} {2010})}\BibitemShut {NoStop}%
\bibitem [{\citenamefont {Hassan}\ and\ \citenamefont
	{S\'en\'echal}(2013)}]{hassan2013}%
\BibitemOpen
\bibfield  {author} {\bibinfo {author} {\bibfnamefont {S.~R.}\ \bibnamefont
		{Hassan}}\ and\ \bibinfo {author} {\bibfnamefont {D.}~\bibnamefont
		{S\'en\'echal}},\ }\href {\doibase 10.1103/PhysRevLett.110.096402} {\bibfield
	{journal} {\bibinfo  {journal} {Phys. Rev. Lett.}\ }\textbf {\bibinfo
		{volume} {110}},\ \bibinfo {pages} {096402} (\bibinfo {year}
	{2013})}\BibitemShut {NoStop}%
\bibitem [{\citenamefont {Liebsch}(2011)}]{LiebschHoneycomb:2011}%
\BibitemOpen
\bibfield  {author} {\bibinfo {author} {\bibfnamefont {A.}~\bibnamefont
		{Liebsch}},\ }\href {\doibase 10.1103/PhysRevB.83.035113} {\bibfield
	{journal} {\bibinfo  {journal} {Phys. Rev. B}\ }\textbf {\bibinfo {volume}
		{83}},\ \bibinfo {pages} {035113} (\bibinfo {year} {2011})}\BibitemShut
{NoStop}%
\bibitem [{\citenamefont {Liebsch}\ and\ \citenamefont
	{Wu}(2013)}]{LiebschWu:2013}%
\BibitemOpen
\bibfield  {author} {\bibinfo {author} {\bibfnamefont {A.}~\bibnamefont
		{Liebsch}}\ and\ \bibinfo {author} {\bibfnamefont {W.}~\bibnamefont {Wu}},\
}\href {\doibase 10.1103/PhysRevB.87.205127} {\bibfield  {journal} {\bibinfo
	{journal} {Phys. Rev. B}\ }\textbf {\bibinfo {volume} {87}},\ \bibinfo
{pages} {205127} (\bibinfo {year} {2013})}\BibitemShut {NoStop}%
\bibitem [{\citenamefont {He}\ and\ \citenamefont
	{Lu}(2012)}]{he_cluster_2012}%
\BibitemOpen
\bibfield  {author} {\bibinfo {author} {\bibfnamefont {R.-Q.}\ \bibnamefont
		{He}}\ and\ \bibinfo {author} {\bibfnamefont {Z.-Y.}\ \bibnamefont {Lu}},\
}\href {\doibase 10.1103/PhysRevB.86.045105} {\bibfield  {journal} {\bibinfo
	{journal} {Physical Review B}\ }\textbf {\bibinfo {volume} {86}},\ \bibinfo
{pages} {045105} (\bibinfo {year} {2012})}\BibitemShut {NoStop}%
\bibitem [{\citenamefont {Seki}\ and\ \citenamefont
	{Ohta}(2013)}]{seki_variational_2013}%
\BibitemOpen
\bibfield  {author} {\bibinfo {author} {\bibfnamefont {K.}~\bibnamefont
		{Seki}}\ and\ \bibinfo {author} {\bibfnamefont {Y.}~\bibnamefont {Ohta}},\
}\href {\doibase 10.3938/jkps.62.2150} {\bibfield  {journal} {\bibinfo
	{journal} {Journal of the Korean Physical Society}\ }\textbf {\bibinfo
	{volume} {62}},\ \bibinfo {pages} {2150} (\bibinfo {year}
{2013})}\BibitemShut {NoStop}%
\bibitem [{Not({\natexlab{a}})}]{Note_1_Arya:2013}%
\BibitemOpen
\href@noop {} {\bibfield  {journal} {\bibinfo  {journal}
		{Studies~\cite{he_cluster_2012,seki_variational_2013} based on the the
			Variational Cluster approximation even find that the spin-liquid phase begins
			at $U=0^+$. However, in the absence of a bath, this approach lacks a metallic
			order parameter, biasing the result towards the insulating phase.}\ }
	({\natexlab{a}})}\BibitemShut {NoStop}%
\bibitem [{\citenamefont {Hohenadler}\ and\ \citenamefont
	{Assaad}(2013)}]{hohenadler_correlation_2013}%
\BibitemOpen
\bibfield  {author} {\bibinfo {author} {\bibfnamefont {M.}~\bibnamefont
		{Hohenadler}}\ and\ \bibinfo {author} {\bibfnamefont {F.~F.}\ \bibnamefont
		{Assaad}},\ }\href {\doibase 10.1088/0953-8984/25/14/143201} {\bibfield
	{journal} {\bibinfo  {journal} {Journal of Physics: Condensed Matter}\
	}\textbf {\bibinfo {volume} {25}},\ \bibinfo {pages} {143201} (\bibinfo
	{year} {2013})}\BibitemShut {NoStop}%
\bibitem [{Not({\natexlab{b}})}]{Note_2_Arya:2013}%
\BibitemOpen
\href@noop {} {\bibfield  {journal} {\bibinfo  {journal} {One cannot however
			completely neglect the short-range spin correlations in determining the
			critical $U/W$ for the Mott transition since single-site Dynamical Mean-Field
			Theory
			~\cite{tran_finite-temperature_2009,jafari_dynamical_2009,ebrahimkhas_exact_2011}
			finds a value much larger than that found on clusters.}\ }
	({\natexlab{b}})}\BibitemShut {NoStop}%
\bibitem [{\citenamefont {{Y.M. Vilk}}\ and\ \citenamefont {{A.-M.S.
			Tremblay}}(1997)}]{Vilk:1997}%
\BibitemOpen
\bibfield  {author} {\bibinfo {author} {\bibnamefont {{Y.M. Vilk}}}\ and\
	\bibinfo {author} {\bibnamefont {{A.-M.S. Tremblay}}},\ }\href {\doibase
	10.1051/jp1:1997135} {\bibfield  {journal} {\bibinfo  {journal} {J. Phys. I
			France}\ }\textbf {\bibinfo {volume} {7}},\ \bibinfo {pages} {1309} (\bibinfo
	{year} {1997})}\BibitemShut {NoStop}%
\bibitem [{\citenamefont {Tremblay}(2011)}]{TremblayMancini:2011}%
\BibitemOpen
\bibfield  {author} {\bibinfo {author} {\bibfnamefont {A.~M.~S.}\
		\bibnamefont {Tremblay}},\ }in\ \href@noop {} {\emph {\bibinfo {booktitle}
		{Strongly Correlated Systems: Theoretical Methods}}},\ \bibinfo {editor}
{edited by\ \bibinfo {editor} {\bibfnamefont {F.}~\bibnamefont {Mancini}}\
	and\ \bibinfo {editor} {\bibfnamefont {A.}~\bibnamefont {Avella}}}\ (\bibinfo
{publisher} {Springer series},\ \bibinfo {year} {2011})\ Chap.~\bibinfo
{chapter} {13}, pp.\ \bibinfo {pages} {409--455}\BibitemShut {NoStop}%
\bibitem [{\citenamefont {Sorella}\ and\ \citenamefont
	{Tosatti}(1992)}]{sorella_semi-metal-insulator_1992}%
\BibitemOpen
\bibfield  {author} {\bibinfo {author} {\bibfnamefont {S.}~\bibnamefont
		{Sorella}}\ and\ \bibinfo {author} {\bibfnamefont {E.}~\bibnamefont
		{Tosatti}},\ }\href {\doibase 10.1209/0295-5075/19/8/007} {\bibfield
	{journal} {\bibinfo  {journal} {{EPL} (Europhysics Letters)}\ }\textbf
	{\bibinfo {volume} {19}},\ \bibinfo {pages} {699} (\bibinfo {year}
	{1992})}\BibitemShut {NoStop}%
\bibitem [{\citenamefont {Martelo}\ \emph {et~al.}(1997)\citenamefont
	{Martelo}, \citenamefont {Dzierzawa}, \citenamefont {Siffert},\ and\
	\citenamefont {Baeriswyl}}]{martelo:1997}%
\BibitemOpen
\bibfield  {author} {\bibinfo {author} {\bibfnamefont {L.}~\bibnamefont
		{Martelo}}, \bibinfo {author} {\bibfnamefont {M.}~\bibnamefont {Dzierzawa}},
	\bibinfo {author} {\bibfnamefont {L.}~\bibnamefont {Siffert}}, \ and\
	\bibinfo {author} {\bibfnamefont {D.}~\bibnamefont {Baeriswyl}},\ }\href@noop
{} {\bibfield  {journal} {\bibinfo  {journal} {Zeitschrift f{\"u}r Physik B
			Condensed Matter}\ }\textbf {\bibinfo {volume} {103}},\ \bibinfo {pages}
	{335} (\bibinfo {year} {1997})}\BibitemShut {NoStop}%
\bibitem [{\citenamefont {Furukawa}(2001)}]{furukawa_antiferromagnetism_2001}%
\BibitemOpen
\bibfield  {author} {\bibinfo {author} {\bibfnamefont {N.}~\bibnamefont
		{Furukawa}},\ }\href {\doibase 10.1143/JPSJ.70.1483} {\bibfield  {journal}
	{\bibinfo  {journal} {Journal of the Physical Society of Japan}\ }\textbf
	{\bibinfo {volume} {70}},\ \bibinfo {pages} {1483} (\bibinfo {year}
	{2001})}\BibitemShut {NoStop}%
\bibitem [{\citenamefont {Assaad}\ and\ \citenamefont
	{Herbut}(2013)}]{pinning_assaad_2013}%
\BibitemOpen
\bibfield  {author} {\bibinfo {author} {\bibfnamefont {F.~F.}\ \bibnamefont
		{Assaad}}\ and\ \bibinfo {author} {\bibfnamefont {I.~F.}\ \bibnamefont
		{Herbut}},\ }\href {\doibase 10.1103/PhysRevX.3.031010} {\bibfield  {journal}
	{\bibinfo  {journal} {Phys. Rev. X}\ }\textbf {\bibinfo {volume} {3}},\
	\bibinfo {pages} {031010} (\bibinfo {year} {2013})}\BibitemShut {NoStop}%
\bibitem [{\citenamefont {Honerkamp}(2008)}]{honerkamp_density_2008}%
\BibitemOpen
\bibfield  {author} {\bibinfo {author} {\bibfnamefont {C.}~\bibnamefont
		{Honerkamp}},\ }\href {\doibase 10.1103/PhysRevLett.100.146404} {\bibfield
	{journal} {\bibinfo  {journal} {Physical Review Letters}\ }\textbf {\bibinfo
		{volume} {100}},\ \bibinfo {pages} {146404} (\bibinfo {year}
	{2008})}\BibitemShut {NoStop}%
\bibitem [{\citenamefont {Raghu}\ \emph {et~al.}(2008)\citenamefont {Raghu},
	\citenamefont {Qi}, \citenamefont {Honerkamp},\ and\ \citenamefont
	{Zhang}}]{raghu_topological_2008}%
\BibitemOpen
\bibfield  {author} {\bibinfo {author} {\bibfnamefont {S.}~\bibnamefont
		{Raghu}}, \bibinfo {author} {\bibfnamefont {X.-L.}\ \bibnamefont {Qi}},
	\bibinfo {author} {\bibfnamefont {C.}~\bibnamefont {Honerkamp}}, \ and\
	\bibinfo {author} {\bibfnamefont {S.-C.}\ \bibnamefont {Zhang}},\ }\href
{\doibase 10.1103/PhysRevLett.100.156401} {\bibfield  {journal} {\bibinfo
		{journal} {Physical Review Letters}\ }\textbf {\bibinfo {volume} {100}},\
	\bibinfo {pages} {156401} (\bibinfo {year} {2008})}\BibitemShut {NoStop}%
\bibitem [{\citenamefont {Vilk}\ \emph {et~al.}(1994)\citenamefont {Vilk},
	\citenamefont {Chen},\ and\ \citenamefont {Tremblay}}]{Vilk:1994}%
\BibitemOpen
\bibfield  {author} {\bibinfo {author} {\bibfnamefont {Y.~M.}\ \bibnamefont
		{Vilk}}, \bibinfo {author} {\bibfnamefont {L.}~\bibnamefont {Chen}}, \ and\
	\bibinfo {author} {\bibfnamefont {A.-M.~S.}\ \bibnamefont {Tremblay}},\
}\href {\doibase 10.1103/PhysRevB.49.13267} {\bibfield  {journal} {\bibinfo
	{journal} {Phys. Rev. B}\ }\textbf {\bibinfo {volume} {49}},\ \bibinfo
{pages} {13267} (\bibinfo {year} {1994})}\BibitemShut {NoStop}%
\bibitem [{\citenamefont {Vilk}\ and\ \citenamefont
	{Tremblay}(1996)}]{Vilk:1996}%
\BibitemOpen
\bibfield  {author} {\bibinfo {author} {\bibfnamefont {Y.~M.}\ \bibnamefont
		{Vilk}}\ and\ \bibinfo {author} {\bibfnamefont {A.-M.~S.}\ \bibnamefont
		{Tremblay}},\ }\href {http://stacks.iop.org/0295-5075/33/i=2/a=159}
{\bibfield  {journal} {\bibinfo  {journal} {EPL (Europhysics Letters)}\
	}\textbf {\bibinfo {volume} {33}},\ \bibinfo {pages} {159} (\bibinfo {year}
	{1996})}\BibitemShut {NoStop}%
\bibitem [{\citenamefont {Allen}\ \emph {et~al.}(2003)\citenamefont {Allen},
	\citenamefont {Tremblay},\ and\ \citenamefont {Vilk}}]{Allen:2003}%
\BibitemOpen
\bibfield  {author} {\bibinfo {author} {\bibfnamefont {S.}~\bibnamefont
		{Allen}}, \bibinfo {author} {\bibfnamefont {A.-M.~S.}\ \bibnamefont
		{Tremblay}}, \ and\ \bibinfo {author} {\bibfnamefont {Y.~M.}\ \bibnamefont
		{Vilk}},\ }in\ \href@noop {} {\emph {\bibinfo {booktitle} {Theoretical
			Methods for Strongly Correlated Electrons}}},\ \bibinfo {editor} {edited by\
	\bibinfo {editor} {\bibfnamefont {D.}~\bibnamefont {S{\'e}n{\'e}chal}},
	\bibinfo {editor} {\bibfnamefont {C.}~\bibnamefont {Bourbonnais}}, \ and\
	\bibinfo {editor} {\bibfnamefont {A.-M.~S.}\ \bibnamefont {Tremblay}}}\
(\bibinfo {year} {2003})\BibitemShut {NoStop}%
\bibitem [{\citenamefont {Bickers}\ and\ \citenamefont
	{Scalapino}(1989)}]{Bickers:1989}%
\BibitemOpen
\bibfield  {author} {\bibinfo {author} {\bibfnamefont {N.~E.}\ \bibnamefont
		{Bickers}}\ and\ \bibinfo {author} {\bibfnamefont {D.~J.}\ \bibnamefont
		{Scalapino}},\ }\href {http://link.aps.org/doi/10.1103/PhysRevLett.62.961}
{\bibfield  {journal} {\bibinfo  {journal} {Ann. Phys. (USA)}\ }\textbf
	{\bibinfo {volume} {193}},\ \bibinfo {pages} {206 } (\bibinfo {year}
	{1989})}\BibitemShut {NoStop}%
\bibitem [{\citenamefont {Bickers}\ and\ \citenamefont
	{White}(1991)}]{Bickers:1991}%
\BibitemOpen
\bibfield  {author} {\bibinfo {author} {\bibfnamefont {N.~E.}\ \bibnamefont
		{Bickers}}\ and\ \bibinfo {author} {\bibfnamefont {S.~R.}\ \bibnamefont
		{White}},\ }\href {\doibase 10.1103/PhysRevB.43.8044} {\bibfield  {journal}
	{\bibinfo  {journal} {Phys. Rev. B}\ }\textbf {\bibinfo {volume} {43}},\
	\bibinfo {pages} {8044} (\bibinfo {year} {1991})}\BibitemShut {NoStop}%
\bibitem [{\citenamefont {Mermin}\ and\ \citenamefont
	{Wagner}(1966)}]{Mermin:1966}%
\BibitemOpen
\bibfield  {author} {\bibinfo {author} {\bibfnamefont {N.~D.}\ \bibnamefont
		{Mermin}}\ and\ \bibinfo {author} {\bibfnamefont {H.}~\bibnamefont
		{Wagner}},\ }\href {\doibase 10.1103/PhysRevLett.17.1133} {\bibfield
	{journal} {\bibinfo  {journal} {Phys. Rev. Lett.}\ }\textbf {\bibinfo
		{volume} {17}},\ \bibinfo {pages} {1133} (\bibinfo {year}
	{1966})}\BibitemShut {NoStop}%
\bibitem [{\citenamefont {Kyung}\ \emph {et~al.}(2004)\citenamefont {Kyung},
	\citenamefont {Hankevych}, \citenamefont {Dar\'e},\ and\ \citenamefont
	{Tremblay}}]{Kyung:2004}%
\BibitemOpen
\bibfield  {author} {\bibinfo {author} {\bibfnamefont {B.}~\bibnamefont
		{Kyung}}, \bibinfo {author} {\bibfnamefont {V.}~\bibnamefont {Hankevych}},
	\bibinfo {author} {\bibfnamefont {A.-M.}\ \bibnamefont {Dar\'e}}, \ and\
	\bibinfo {author} {\bibfnamefont {A.-M.~S.}\ \bibnamefont {Tremblay}},\
}\href {\doibase 10.1103/PhysRevLett.93.147004} {\bibfield  {journal}
{\bibinfo  {journal} {Phys. Rev. Lett.}\ }\textbf {\bibinfo {volume} {93}},\
\bibinfo {pages} {147004} (\bibinfo {year} {2004})}\BibitemShut {NoStop}%
\bibitem [{\citenamefont {Hassan}\ \emph {et~al.}(2008)\citenamefont {Hassan},
	\citenamefont {Davoudi}, \citenamefont {Kyung},\ and\ \citenamefont
	{Tremblay}}]{Hassan:2008}%
\BibitemOpen
\bibfield  {author} {\bibinfo {author} {\bibfnamefont {S.~R.}\ \bibnamefont
		{Hassan}}, \bibinfo {author} {\bibfnamefont {B.}~\bibnamefont {Davoudi}},
	\bibinfo {author} {\bibfnamefont {B.}~\bibnamefont {Kyung}}, \ and\ \bibinfo
	{author} {\bibfnamefont {A.-M.~S.}\ \bibnamefont {Tremblay}},\ }\href
{\doibase 10.1103/PhysRevB.77.094501} {\bibfield  {journal} {\bibinfo
		{journal} {Phys. Rev. B}\ }\textbf {\bibinfo {volume} {77}},\ \bibinfo
	{pages} {094501} (\bibinfo {year} {2008})}\BibitemShut {NoStop}%
\bibitem [{\citenamefont {Kyung}\ \emph {et~al.}(2003)\citenamefont {Kyung},
	\citenamefont {Landry},\ and\ \citenamefont {Tremblay}}]{Kyung:2003}%
\BibitemOpen
\bibfield  {author} {\bibinfo {author} {\bibfnamefont {B.}~\bibnamefont
		{Kyung}}, \bibinfo {author} {\bibfnamefont {J.-S.}\ \bibnamefont {Landry}}, \
	and\ \bibinfo {author} {\bibfnamefont {A.~M.~S.}\ \bibnamefont {Tremblay}},\
}\href {http://link.aps.org/doi/10.1103/PhysRevB.68.174502} {\bibfield
{journal} {\bibinfo  {journal} {Phys. Rev. B}\ }\textbf {\bibinfo {volume}
	{68}},\ \bibinfo {pages} {174502} (\bibinfo {year} {2003})}\BibitemShut
{NoStop}%
\bibitem [{\citenamefont {Allen}\ and\ \citenamefont
	{Tremblay}(2001)}]{Allen:2001}%
\BibitemOpen
\bibfield  {author} {\bibinfo {author} {\bibfnamefont {S.}~\bibnamefont
		{Allen}}\ and\ \bibinfo {author} {\bibfnamefont {A.-M.~S.}\ \bibnamefont
		{Tremblay}},\ }\href@noop {} {\bibfield  {journal} {\bibinfo  {journal}
		{Phys. Rev. B}\ }\textbf {\bibinfo {volume} {64}},\ \bibinfo {pages} {075115
	} (\bibinfo {year} {2001})}\BibitemShut {NoStop}%
\bibitem [{\citenamefont {Davoudi}\ and\ \citenamefont
	{Tremblay}(2006)}]{davoudi:2006}%
\BibitemOpen
\bibfield  {author} {\bibinfo {author} {\bibfnamefont {B.}~\bibnamefont
		{Davoudi}}\ and\ \bibinfo {author} {\bibfnamefont {A.-M.~S.}\ \bibnamefont
		{Tremblay}},\ }\href {\doibase 10.1103/PhysRevB.74.035113} {\bibfield
	{journal} {\bibinfo  {journal} {Phys. Rev. B}\ }\textbf {\bibinfo {volume}
		{74}},\ \bibinfo {pages} {035113} (\bibinfo {year} {2006})}\BibitemShut
{NoStop}%
\bibitem [{\citenamefont {Davoudi}\ and\ \citenamefont
	{Tremblay}(2007)}]{davoudi:2007}%
\BibitemOpen
\bibfield  {author} {\bibinfo {author} {\bibfnamefont {B.}~\bibnamefont
		{Davoudi}}\ and\ \bibinfo {author} {\bibfnamefont {A.-M.~S.}\ \bibnamefont
		{Tremblay}},\ }\href {\doibase 10.1103/PhysRevB.76.085115} {\bibfield
	{journal} {\bibinfo  {journal} {Phys. Rev. B}\ }\textbf {\bibinfo {volume}
		{76}},\ \bibinfo {pages} {085115} (\bibinfo {year} {2007})}\BibitemShut
{NoStop}%
\bibitem [{\citenamefont {Davoudi}\ \emph {et~al.}(2008)\citenamefont
	{Davoudi}, \citenamefont {Hassan},\ and\ \citenamefont
	{Tremblay}}]{davoudi:2008}%
\BibitemOpen
\bibfield  {author} {\bibinfo {author} {\bibfnamefont {B.}~\bibnamefont
		{Davoudi}}, \bibinfo {author} {\bibfnamefont {S.~R.}\ \bibnamefont {Hassan}},
	\ and\ \bibinfo {author} {\bibfnamefont {A.-M.~S.}\ \bibnamefont
		{Tremblay}},\ }\href {\doibase 10.1103/PhysRevB.77.214408} {\bibfield
	{journal} {\bibinfo  {journal} {Phys. Rev. B}\ }\textbf {\bibinfo {volume}
		{77}},\ \bibinfo {pages} {214408} (\bibinfo {year} {2008})}\BibitemShut
{NoStop}%
\bibitem [{\citenamefont {Moukouri}\ \emph {et~al.}(2000)\citenamefont
	{Moukouri}, \citenamefont {Allen}, \citenamefont {Lemay}, \citenamefont
	{Kyung}, \citenamefont {Poulin}, \citenamefont {Vilk},\ and\ \citenamefont
	{Tremblay}}]{Moukouri:2000}%
\BibitemOpen
\bibfield  {author} {\bibinfo {author} {\bibfnamefont {S.}~\bibnamefont
		{Moukouri}}, \bibinfo {author} {\bibfnamefont {S.}~\bibnamefont {Allen}},
	\bibinfo {author} {\bibfnamefont {F.}~\bibnamefont {Lemay}}, \bibinfo
	{author} {\bibfnamefont {B.}~\bibnamefont {Kyung}}, \bibinfo {author}
	{\bibfnamefont {D.}~\bibnamefont {Poulin}}, \bibinfo {author} {\bibfnamefont
		{Y.~M.}\ \bibnamefont {Vilk}}, \ and\ \bibinfo {author} {\bibfnamefont
		{A.-M.~S.}\ \bibnamefont {Tremblay}},\ }\href {\doibase
	10.1103/PhysRevB.61.7887} {\bibfield  {journal} {\bibinfo  {journal} {Phys.
			Rev. B}\ }\textbf {\bibinfo {volume} {61}},\ \bibinfo {pages} {7887}
	(\bibinfo {year} {2000})}\BibitemShut {NoStop}%
\bibitem [{\citenamefont {Dar\'e}\ \emph {et~al.}(1996)\citenamefont {Dar\'e},
	\citenamefont {Vilk},\ and\ \citenamefont {Tremblay}}]{Dare:1996}%
\BibitemOpen
\bibfield  {author} {\bibinfo {author} {\bibfnamefont {A.-M.}\ \bibnamefont
		{Dar\'e}}, \bibinfo {author} {\bibfnamefont {Y.~M.}\ \bibnamefont {Vilk}}, \
	and\ \bibinfo {author} {\bibfnamefont {A.~M.~S.}\ \bibnamefont {Tremblay}},\
}\href {\doibase 10.1103/PhysRevB.53.14236} {\bibfield  {journal} {\bibinfo
	{journal} {Phys. Rev. B}\ }\textbf {\bibinfo {volume} {53}},\ \bibinfo
{pages} {14236} (\bibinfo {year} {1996})}\BibitemShut {NoStop}%
\bibitem [{\citenamefont {Herbut}(2006)}]{herbut_interactions_2006}%
\BibitemOpen
\bibfield  {author} {\bibinfo {author} {\bibfnamefont {I.~F.}\ \bibnamefont
		{Herbut}},\ }\href {\doibase 10.1103/PhysRevLett.97.146401} {\bibfield
	{journal} {\bibinfo  {journal} {Physical Review Letters}\ }\textbf {\bibinfo
		{volume} {97}},\ \bibinfo {pages} {146401} (\bibinfo {year}
	{2006})}\BibitemShut {NoStop}%
\bibitem [{\citenamefont {Bergeron}\ \emph {et~al.}(2011)\citenamefont
	{Bergeron}, \citenamefont {Hankevych}, \citenamefont {Kyung},\ and\
	\citenamefont {Tremblay}}]{Bergeron:2011}%
\BibitemOpen
\bibfield  {author} {\bibinfo {author} {\bibfnamefont {D.}~\bibnamefont
		{Bergeron}}, \bibinfo {author} {\bibfnamefont {V.}~\bibnamefont {Hankevych}},
	\bibinfo {author} {\bibfnamefont {B.}~\bibnamefont {Kyung}}, \ and\ \bibinfo
	{author} {\bibfnamefont {A.-M.~S.}\ \bibnamefont {Tremblay}},\ }\href
{\doibase 10.1103/PhysRevB.84.085128} {\bibfield  {journal} {\bibinfo
		{journal} {Phys. Rev. B}\ }\textbf {\bibinfo {volume} {84}},\ \bibinfo
	{pages} {085128} (\bibinfo {year} {2011})}\BibitemShut {NoStop}%
\bibitem [{\citenamefont {Kanamori}(1963)}]{kanamori_electron_1963}%
\BibitemOpen
\bibfield  {author} {\bibinfo {author} {\bibfnamefont {J.}~\bibnamefont
		{Kanamori}},\ }\href {\doibase 10.1143/PTP.30.275} {\bibfield  {journal}
	{\bibinfo  {journal} {Progress of Theoretical Physics}\ }\textbf {\bibinfo
		{volume} {30}},\ \bibinfo {pages} {275} (\bibinfo {year} {1963})}\BibitemShut
{NoStop}%
\bibitem [{\citenamefont {Vilk}\ and\ \citenamefont
	{Tremblay}(1995)}]{Vilk:1995}%
\BibitemOpen
\bibfield  {author} {\bibinfo {author} {\bibfnamefont {Y.~M.}\ \bibnamefont
		{Vilk}}\ and\ \bibinfo {author} {\bibfnamefont {A.-M.~S.}\ \bibnamefont
		{Tremblay}},\ }\href@noop {} {\bibfield  {journal} {\bibinfo  {journal} {J.
			Phys. Chem. Solids (UK)}\ }\textbf {\bibinfo {volume} {56}},\ \bibinfo
	{pages} {1769 } (\bibinfo {year} {1995})}\BibitemShut {NoStop}%
\bibitem [{\citenamefont {Miyahara}\ \emph {et~al.}(2013)\citenamefont
	{Miyahara}, \citenamefont {Arita},\ and\ \citenamefont
	{Ikeda}}]{AritaTPSC:2013}%
\BibitemOpen
\bibfield  {author} {\bibinfo {author} {\bibfnamefont {H.}~\bibnamefont
		{Miyahara}}, \bibinfo {author} {\bibfnamefont {R.}~\bibnamefont {Arita}}, \
	and\ \bibinfo {author} {\bibfnamefont {H.}~\bibnamefont {Ikeda}},\ }\href
{\doibase 10.1103/PhysRevB.87.045113} {\bibfield  {journal} {\bibinfo
		{journal} {Phys. Rev. B}\ }\textbf {\bibinfo {volume} {87}},\ \bibinfo
	{pages} {045113} (\bibinfo {year} {2013})}\BibitemShut {NoStop}%
\end{thebibliography}

%

\end{document}